\documentclass[aip,graphicx]{revtex4-1}
\usepackage{amsmath}
\usepackage{graphicx}
\usepackage{color}
\draft 

\begin{document}

\title{Coupling Polyatomic Molecules to Lossy Nanocavities: Lindblad versus Schr\"odinger description}

\author{Csaba F\'abri}
\email{ficsaba@staff.elte.hu}
\affiliation{HUN-REN--ELTE Complex Chemical Systems Research Group, P.O. Box 32, H-1518 Budapest 112, Hungary}
\affiliation{Department of Theoretical Physics, University of Debrecen, P.O. Box 400, H-4002 Debrecen, Hungary}

\author{Attila G. Cs\'asz\'ar}
\affiliation{HUN-REN--ELTE Complex Chemical Systems Research Group, P.O. Box 32, H-1518 Budapest 112, Hungary}
\affiliation{Laboratory of Molecular Structure and Dynamics, Institute of Chemistry,
ELTE E\"otv\"os Lor\'and University, H-1117 Budapest, 
P\'azm\'any P\'eter s\'et\'any 1/A, Hungary}

\author{G\'abor J. Hal\'asz}
\affiliation{Department of Information Technology, University of Debrecen, P.O. Box 400, H-4002 Debrecen, Hungary}

\author{Lorenz S. Cederbaum}
\affiliation{Theoretische Chemie, Physikalisch-Chemisches Institut, Universit\"at Heidelberg, D-69120 Heidelberg, Germany}

\author{\'Agnes Vib\'ok}
\email{vibok@phys.unideb.hu}
\affiliation{Department of Theoretical Physics, University of Debrecen, P.O. Box 400, H-4002 Debrecen, Hungary}
\affiliation{ELI-ALPS, ELI-HU Non-Profit Ltd, H-6720 Szeged, Dugonics t\'er 13, Hungary}

\date{\today}

\begin{abstract}
The use of cavities to impact molecular structure and dynamics has become popular. 
As cavities, in particular plasmonic nanocavities, are lossy and the lifetime of their modes 
can be very short, their lossy nature must be incorporated into the calculations.
The Lindblad master equation is commonly considered as an appropriate tool to describe this 
lossy nature.  This approach requires the dynamics of the density operator and is thus 
substantially more costly than approaches employing the Schr\"odinger equation for the quantum
wave function when several or many nuclear degrees of freedom are involved. In this work we 
compare numerically the Lindblad and Schr\"odinger descriptions discussed in the literature 
for a molecular example where the cavity is pumped by a laser. The laser and cavity properties
are varied over a range of parameters. It is found that the Schr\"odinger description adequately
describes the dynamics of the polaritons and emission signal as long as the laser intensity is
moderate and the pump time is not much longer than the lifetime of the cavity mode. Otherwise, 
it is demonstrated that the Schr\"odinger description gradually fails. 
We also show that the failure of the Schr\"odinger description can often be
remedied by renormalizing the wave function at every step of the time propagation.
The results are discussed and analyzed.
\end{abstract}
\pacs{}

\maketitle 

\section{Introduction}
\label{sec:intro}

Polyatomic quantum emitters offer a high degree of complexity as electron-photon
interactions are strongly influenced by the coupling of the electrons
to nuclear vibrational and rotational motions. The confined photonic
mode of the cavity can resonantly couple to the transition dipole
of the molecule, giving rise to mixed polaritonic states carrying both
photonic and excitonic features.\cite{15ToBa} A vast number of
experimental \cite{12HuScGe,16Ebbesen,16ThGeSh,16VeGeCh,16ZhChWa,16ChNiBe,19OjChDe,19RoShEr,19ThLeNa,19VeThNa}
and theoretical \cite{15GaGaFe,16KoBeMu,18FeGaGa,18FlRiNa,18RiMaDu,18RuTaFl,18SzHaCs_2,18TrPeSa,18Vendrell,19CsKoHa,19CsViHa,19KeZh,19MaHu,19TrSa,20FaLaHa_2,20FrCoPe,20HeOw,20MaMoHu,20SiPiGa,20SiScRu,20SzHaVi,20TaMaZh,21Cederbaum,21CeKu,21FaHaCe,21FaHaCe_2,21FiSa,21GaFrCi,21LiNiSu,22CuNi,22FrGaFe,22LiCuSu,22ReSoGe,22WePuSc,23Fabri,23FaShDo,23FrCo,23MaTaWe,23PeKoSt,23ScKo,23ScSiRu}
works have demonstrated that hybrid light-matter polaritons
can dramatically modify and even control static and dynamic features
of matter.
Interesting applications have reported the enhancement
or suppression of photochemical processes, reaction rates\cite{16GaGaFe,19MaHu,19CaRiZh,20FrCoPe,20FeFrSc,20DaKo_2}
and isomerization,\cite{18FrGrCo,20FrGrPe} the modification of charge \cite{19ScRuAp} or electron \cite{19SeNi,20MaKrHu,21WePuSc}
transfer between molecules, as well as the emergence of light-induced nonadiabatic
effects, \cite{16KoBeMu,18FeGaGa,18SzHaCs_2,18Vendrell,19CsKoHa,19CsViHa,19PeJuYu,19UlGoVe,20FaLaHa_2,20GuMu,20GuMu_2,20SzHaVi,21FaHaCe,21FaHaCe_2,21FaMaHu,21SzBaHa,22CsVeHa,22FaHaCe,22FaHaVi,22FiSa,23ScSiRu}
and so forth.

To successfully modify and control molecular properties one has to reach
the strong light-matter coupling regime. This can only be achieved
when the coupling strength between light and matter excitations becomes
larger than the dissipation rates of the cavity mode and molecular excited states, which can be easily realized for a macroscopic
number of molecules coupled to an optical or plasmonic cavity mode. As for
single-molecule experiments, the situation is more complicated.\cite{16ChNiBe}
In this case, strong subwavelength-size confinement is required to reach the
strong-coupling limit at room temperature with typical transition dipole moment values
of organic molecules, which is feasible only in plasmonic nanocavities.
The latter, however, may possess a very lossy nature, because a relevant part of the
energy of the nanocavity mode is stored in the kinetic energy of electrons in the metal.
The lifetime of plasmonic nanocavity modes is very short, typically on the order of
a few tens of femtoseconds, which must be adequately addressed in numerical
simulations.

In the present work, we investigate the quantum dynamics of a single
polyatomic molecule interacting with a lossy plasmonic nanocavity in the strong coupling regime.
Generally, the photon loss in such a quantum system can be described by coupling
the cavity-molecule system to a dissipative Markovian environment, leading to the
Lindblad master equation formalism.\cite{20Manzano,20DaKo_2,20SiPiGa,21ToFe,22FaHaCe,22FaHaVi}
In this situation, the temporal evolution of the system can be obtained by propagating
the density matrix in time, according to the appropriate
Lindblad master equation. However, cavity loss can be incorporated in
the time-dependent Schr\"odinger equation (TDSE) framework as well.\cite{20UlVe,20AnSuVa}
This is an alternative and frequently
used description of how a photon can escape from a cavity,
but the adequacy of this description needs to be explored. 
If quantum dynamics is restricted to a certain excitation manifold (such as the
singly-excited subspace, that is, ground-state molecule  with one photon and excited-state molecule with zero photons), dissipative effects can be described
by a non-Hermitian formalism in which the Hamiltonian of the system
is augmented with an imaginary term. The effective non-Hermitian Hamiltonian
obtained in this manner implicitly includes the cavity leakage which is then accounted
for by a loss of norm of the nuclear wave packet during time propagation using the TDSE. 

In this paper, the performance of the non-Hermitian TDSE method is
compared to that of the Lindblad-master-equation approach.
To carry out a thorough comparison of the two methods, different cavity and laser pumping
parameters are applied. As a showcase system, the four-atomic formaldehyde (H$_2$CO) molecule
is applied. It is demonstrated that if the lower (1LP) and upper
(1UP) polaritonic states in the singly-excited subspace are driven with a pump pulse,
the TDSE with a non-Hermitian Hamiltonian works very well in many
cases as long as excited-state populations and cavity emission are of primary interest. Therefore,
it is worth examining the range of validity of the Schr\"odinger-based
description, also due to its more favorable computational cost. The Lindblad-master-equation
approach is known to be computationally more demanding than the TDSE, as the state of the system
is represented by a density matrix (instead of a state vector) and matrix-matrix multiplications
have to be performed during the time evolution.  Another important advantage of the non-Hermitian
TDSE is  that it enables the inclusion of molecular rotational degrees of freedom in the 
quantum-dynamical description \cite{15HaViCe,13HaViMo} due to its lower computational cost 
compared to the Lindblad-based description. Thus, the non-Hermitian TDSE offers a correct 
treatment of light-induced nonadiabatic properties of small diatomic molecules in a 
lossy nanocavity.

However, in several situations, results of the Lindblad and non-Hermitian TDSE approaches
differ significantly. For example, the ground-state Schr\"odinger and
Lindblad populations are always different. It is also shown that the
Schr\"odinger and Lindblad results can deviate substantially if the lifetime of the
cavity mode is short relative to the length of the pump pulse. Moreover,
excitation above the singly-excited subspace can also lead to visible
differences between the two approaches.
We will also demonstrate that these discrepancies between the Lindblad
and non-Hermitian TDSE approaches can often be remedied by setting the norm of the wave
function to one at each step of the time propagation in case of the non-Hermitian TDSE
(renormalized TDSE method).

A similar comparison of the Lindblad and non-Hermitian TDSE methods
has recently been presented in Ref. \onlinecite{20CoOtGr} where the non-Hermitian TDSE 
approach has been found to often provide adequate results for quantum dots coupled to a
plasmonic mode.
Another relevant work is Ref. \onlinecite{22McFo} where
a non-Hermitian Schr\"odinger equation was employed to develop a
cavity quantum electrodynamics model for dissipative photonic modes using the 
configuration interaction singles method.
Finally, it is worth noting that a compromise between the Lindblad and
non-Hermitian TDSE approaches could be the stochastic Schr\"odinger method based on
the Monte Carlo wave packet formalism.\cite{93MoCaDa,22MaGaBi,22MaGaBi_2}

The structure of the paper is as follows. Section \ref{sec:theory} provides the
(i) theoretical description of the lossy cavity-molecule system, 
(ii) Lindblad-master-equation approach, 
(iii) non-Hermitian TDSE approach, and 
(iv) a brief summary of the computational model used for the H$_2$CO molecule.
Section \ref{sec:results} presents and discusses the results of the quantum-dynamical
computations, while conclusions and outlook are provided in Section \ref{sec:conclusions}.
In addition, results with the renormalized TDSE method, density matrix purities
and results with a special initial state are given in Appendix \ref{sec:appendixA} and
\ref{sec:appendixB}.

\section{Theory and computational protocol}
\label{sec:theory}

A single molecule interacting with a cavity mode can be described by the
Hamiltonian \cite{04CoDuGr}
\begin{equation}
        \hat{H}_\textrm{cm} = \hat{H}_0 + \hbar \omega_\textrm{c} \hat{a}^\dag \hat{a} - g \hat{\vec{\mu}} \vec{e} (\hat{a}^\dag + \hat{a}) +
            \frac{g^2}{\hbar \omega_\textrm{c}} (\hat{\vec{\mu}} \vec{e})^2
   \label{eq:Hcm}
\end{equation}
where $\hat{H}_0$ is the Hamiltonian of the isolated (field-free) molecule, $\omega_\textrm{c}$ 
denotes the cavity-mode angular frequency, $\hat{a}^\dag$ and $\hat{a}$ are creation and annihilation operators of the cavity mode, $\hat{\vec{\mu}}$ is the electric dipole moment operator of the
molecule and $\vec{e}$ corresponds to the cavity field polarization vector.
The coupling between the molecule and the cavity mode is described by
the coupling strength parameter $g = \sqrt{\frac{\hbar \omega_\textrm{c}}{2 \epsilon_0 V}}$ 
where $\epsilon_0$ and $V$ are the permittivity and quantization volume of the cavity, respectively.
The last term of $\hat{H}_\textrm{cm}$ is the so-called dipole self-energy.\cite{18RoWeRu,20ScRuRo,20MaMoHu,21TrSa,22FrGaFe,23SiScOb,23ScSiRu}
In this work, we consider a single molecule coupled to a plasmonic cavity mode
and omit the dipole self-energy (DSE) term (see Refs. \onlinecite{22FrGaFe,19GaClGa,21FeFeGa}
for further explanation).

In one takes into account two molecular electronic states (X and A), $\hat{H}_\textrm{cm}$ can be recast as
\begin{equation}
    \resizebox{0.9\textwidth}{!}{$\hat{H}_\textrm{cm}  = 
         \begin{bmatrix}
            \hat{T} + V_\textrm{X} & 0 & W_{\textrm{X}}^{(1)} & W_{\textrm{XA}}^{(1)} & 0 & 0 & \dots \\
            0 & \hat{T} + V_\textrm{A} & W_{\textrm{XA}}^{(1)} & W_{\textrm{A}}^{(1)} & 0 & 0 & \dots \\
            W_{\textrm{X}}^{(1)} & W_{\textrm{XA}}^{(1)} & \hat{T} + V_\textrm{X} + \hbar\omega_\textrm{c} & 0 & W_{\textrm{X}}^{(2)} & W_{\textrm{XA}}^{(2)} &\dots \\
            W_{\textrm{XA}}^{(1)} & W_{\textrm{A}}^{(1)} & 0 &\hat{T} + V_\textrm{A} + \hbar\omega_\textrm{c} & W_{\textrm{XA}}^{(2)} & W_{\textrm{A}}^{(2)} & \dots \\
            0 & 0 & W_{\textrm{X}}^{(2)} & W_{\textrm{XA}}^{(2)} & \hat{T} + V_\textrm{X} + 2\hbar\omega_\textrm{c} & 0 &\dots \\
            0 & 0 & W_{\textrm{XA}}^{(2)} & W_{\textrm{A}}^{(2)} & 0 &\hat{T} + V_\textrm{A} + 2\hbar\omega_\textrm{c} & \dots \\
            \vdots & \vdots & \vdots & \vdots & \vdots & \vdots & \ddots 
        \end{bmatrix}$}
    \label{eq:cavity_H}
\end{equation}
where $\hat{T}$ is the nuclear (rotational-vibrational) kinetic energy operator, 
while $V_\textrm{X}$ and $V_\textrm{A}$ are the electronic ground-state and excited-state
potential energy surfaces (PESs) of the molecule.
The cavity-molecule coupling is described by the operators $W_\alpha^{(n)} = -g \sqrt{n} \mu_\alpha$
($\alpha = \textrm{X}, \textrm{A}$) and $W_{\textrm{XA}}^{(n)} = -g \sqrt{n} \mu_\textrm{XA}$
where Fock states of the cavity mode are labeled by the quantum number $n = 0,1,2,\dots$.
Projections of the permanent (PDM) and transition (TDM) dipole moments along $\vec{e}$ are
denoted by $\mu_\alpha$ ($\alpha = \textrm{X}, \textrm{A}$) and $\mu_\textrm{XA}$, respectively.
The Hamiltonian $\hat{H}_\textrm{cm}$ of Eq. \eqref{eq:Hcm} corresponds to the
so-called diabatic representation. Polaritonic (adiabatic) PESs can be obtained as
eigenvalues of the potential energy part of $\hat{H}_\textrm{cm}$ at each nuclear configuration.
Under strong coupling, the ground (lowest) polaritonic state essentially equals
$|\textrm{X}0\rangle$ (electronic ground state with zero photons), while the
lower (1LP) and upper (1UP) polaritonic states can be well approximated as the
linear combinations of $|\textrm{X}1\rangle$ and $|\textrm{A}0\rangle$ (singly-excited subspace).

$\hat{H}_\textrm{cm}$ of Eq. \eqref{eq:Hcm} clearly assumes an infinite lifetime for field excitations.
However, the highly lossy nature of plasmonic nanocavities requires that finite photon lifetimes
are taken into account. As already outlined in Section \ref{sec:intro}, this can be
achieved by the Lindblad master equation\cite{20Manzano}
\begin{equation}
\frac{\partial \hat{\rho}}{\partial t} =
		-\frac{\textrm{i}}{\hbar} [\hat{H},\hat{\rho}] + \gamma_\textrm{c} \hat{a} \hat{\rho} \hat{a}^\dag -
		 \frac{\gamma_\textrm{c}}{2} ( \hat{\rho} \hat{N} + \hat{N} \hat{\rho} )
	\label{eq:Lindblad}
\end{equation}
where $\hat{\rho}$ denotes the density operator and $\hat{N} = \hat{a}^\dag \hat{a}$ is the photon
number operator associated with the cavity mode. Moreover, $\gamma_\textrm{c}$
specifies the cavity decay rate which is equivalent to a lifetime of $\tau = 1/\gamma_\textrm{c}$.
In Eq. \eqref{eq:Lindblad}, $\hat{H}_\textrm{cm}$ is supplemented with a term which describes
the interaction of the cavity mode with a pump laser pulse, that is,
\begin{equation}
        \hat{H} = \hat{H}_\textrm{cm} - \mu_\textrm{c} E(t) (\hat{a}^\dag+\hat{a})
  \label{eq:fullH}
\end{equation}
where $\mu_\textrm{c} = 1.0 ~ \textrm{au}$ (effective dipole moment of the cavity mode) and
the pump pulse is specified by
$E(t) = E_0 \sin^2(\pi t / T) \cos(\omega t)$ for $0 \le t \le T$ and $E(t) \equiv 0$ otherwise.
$E_0$, $T$ and $\omega$ denote the amplitude, length and carrier angular frequency of
the pump pulse, respectively. $E_0$ can be converted into laser intensity by the formula
$I = c \epsilon_0 E_0^2 /2$ ($c$ is the speed of light in vacuum and $\epsilon_0$ is
the vacuum permittivity). Following Ref. \onlinecite{20SiPiGa}, we take the radiative emission
rate $E_\textrm{R}$ to be proportional to the expectation value of $\hat{N}$, which yields
$E_\textrm{R} \sim \textrm{Tr} (\hat{\rho} \hat{N}) = N(t)$.
As the lifetime of excited electronic states typically exceeds
the cavity lifetimes applied in the current work, we assume an infinite lifetime for the
excited electronic state.

As explained in Ref. \onlinecite{95ViNi}, dissipation during the time evolution of an open quantum system can be described by the non-Hermitian time-dependent Schr\"odinger equation (TDSE).
In our case, the TDSE takes the form
\begin{equation}
\textrm{i} \hbar \frac{\partial |\Psi\rangle}{\partial t} =
    \Bigl( \hat{H} - \textrm{i} \frac{\gamma_\textrm{c}}{2} \hat{N} \Bigr) |\Psi\rangle
	\label{eq:Schrodinger}
\end{equation}
where the Hamiltonian $\hat{H} - \textrm{i} \frac{\gamma_\textrm{c}}{2} \hat{N}$ is
clearly non-Hermitian. 
It is also shown in Ref. \onlinecite{95ViNi} that the Lindblad master equation without the term
$\gamma_\textrm{c} \hat{a} \hat{\rho} \hat{a}^\dag$
is equivalent to the TDSE of Eq. \eqref{eq:Schrodinger}.
The term $\gamma_\textrm{c} \hat{a} \hat{\rho} \hat{a}^\dag$ induces incoherent transitions $|\alpha (n+1)\rangle \rightarrow |\alpha n\rangle$ ($\alpha$ labels
molecular electronic states X and A).
This way, population losses in the states $|\alpha n\rangle$ with $n>0$ are compensated and the space integral of the density is conserved by the Lindblad master equation.
Obviously, if one employs the non-Hermitian TDSE, this is not the case and the norm of the wave function decreases over time due to the term $-\textrm{i} \frac{\gamma_\textrm{c}}{2} \hat{N}$ in Eq. \eqref{eq:Schrodinger}.
As we will see, one way of improving the non-Hermitian TDSE method is to set
the norm of the wave function to one at each time step during the time propagation, which
is also applied by the stochastic Schr\"odinger method.\cite{93MoCaDa}
The time evolution can be expressed in terms of the state vector $|\Psi\rangle$
instead of $\hat{\rho}$ if the TDSE is used, which significantly reduces the cost of time
propagation in particular if several or many nuclear degrees of freedom are involved.
Consequently, knowing where the Schr\"odinger approach is a good approximation is 
extremely relevant.

Next, the computational model used for the formaldehyde (H$_{2}$CO) molecule is described briefly.
We consider the two singlet electronic states $\textrm{S}_0 ~ (\tilde{\textrm{X}} ~ ^1\textrm{A}_1)$
and $\textrm{S}_1 ~ (\tilde{\textrm{A}} ~ ^1\textrm{A}_2)$ of H$_{2}$CO and treat the
$\nu_2$ (C=O stretch) and $\nu_4$ (out-of-plane) vibrational modes (2D($\nu_2$,$\nu_4$)
vibrational model) in all computations. We stress that the 2D($\nu_2$,$\nu_4$) model has been
shown to provide a physically correct description of the X$\rightarrow$A electronic spectrum of
H$_{2}$CO.\cite{20FaLaHa_2,21FaHaCe}
In addition, the three rotational degrees of freedom are omitted and the molecular
orientation is kept fixed with respect to the polarization vector ($\vec{e}$) of the cavity field.
In the 2D($\nu_2$,$\nu_4$) model, the PDMs are orthogonal to the TDM. Therefore, as $\vec{e}$ is chosen to be parallel with the TDM in the
current work, the PDMs are orthogonal to $\vec{e}$
and the terms $W_\alpha^{(n)}$  vanish in Eq. \eqref{eq:cavity_H}.
We refer to Refs. \onlinecite{20FaLaHa,20FaLaHa_2,21FaHaCe_2,22FaHaVi,22FaHaCe}
for further details of the 2D($\nu_2$,$\nu_4$) vibrational model.

In case of the Lindblad master equation, the state of the coupled cavity-molecule system is
given by the density operator $\hat{\rho}$. In order to facilitate numerical computations,
$\hat{\rho}$ is represented in the basis $|\alpha i n\rangle$ where $\alpha=\textrm{X},\textrm{A}$
labels molecular electronic states, $i$ denotes two-dimensional direct-product discrete variable representation (DVR)\cite{65HaEnGw,00LiCa} basis functions and $n = 0,1,2,3$.
In the current work, an equidistant Fourier DVR\cite{70Meyer} has been employed for
both the $\nu_2$ and $\nu_4$ vibrational modes.
For the TDSE, one can again utilize the basis $|\alpha i n\rangle$ to
represent $|\Psi\rangle$. The Lindblad master equation and the TDSE are integrated
with the explicit third-order Adams--Bashforth method.\cite{03SuMa}

\begin{figure}
\includegraphics[width=0.375\textwidth]{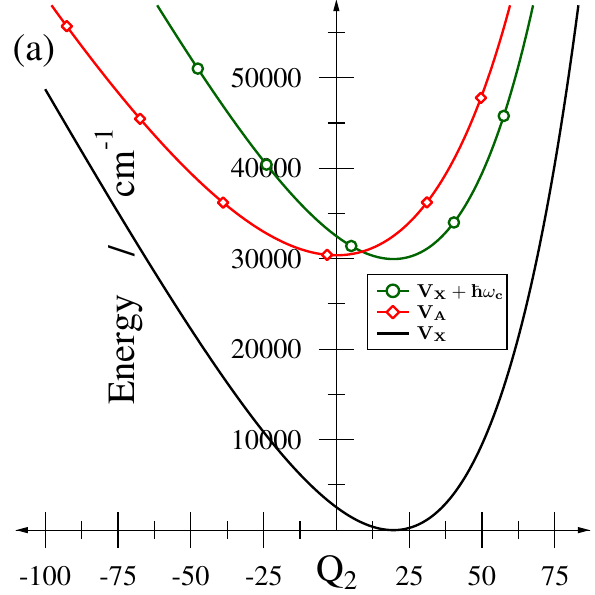}
\includegraphics[width=0.615\textwidth]{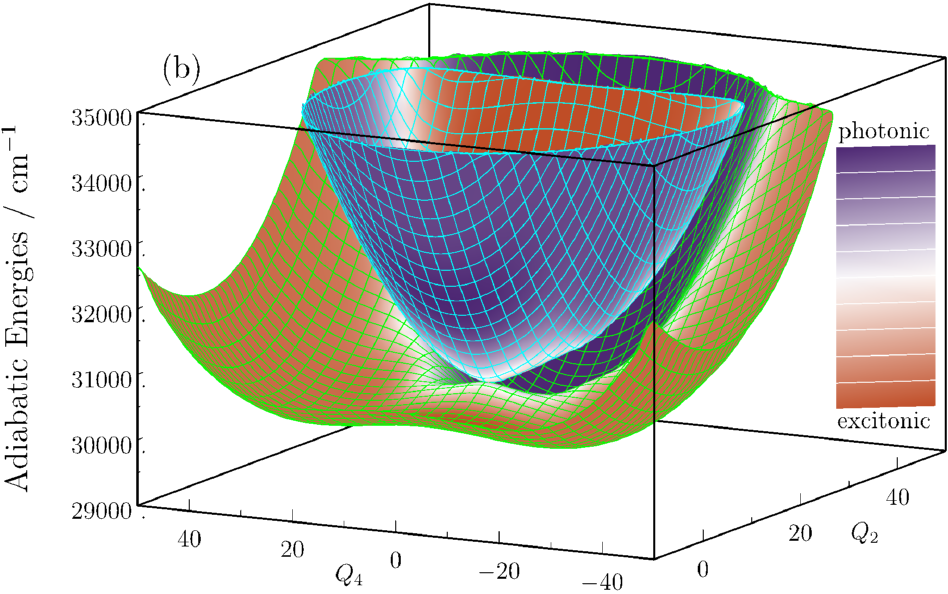}
\caption{\label{fig:PES}
(a) Diabatic potentials ($V_\textrm{X}$, $V_\textrm{A}$ and $V_\textrm{X}+\hbar\omega_{\textrm{c}}$) as a function of the C=O stretch normal coordinate ($Q_2$) (the out-of-plane normal coordinate is fixed at $Q_4=0$). The cavity wavenumber equals $\omega_{\textrm{c}}=29957.2~\textrm{cm}^{-1}$.
(b) Two-dimensional lower (1LP) and upper (1UP) polaritonic potential energy surfaces
(singly-excited subspace). The cavity wavenumber and coupling strength are set to 
$\omega_{\textrm{c}}=29957.2~\textrm{cm}^{-1}$ and $g=0.02~\textrm{au}$, respectively. 
The character of the polaritonic surfaces is indicated by a purple-orange colormap 
(purple: photonic (ground-state molecule with one photon), 
orange: excitonic (excited-state molecule with zero photons)).
}
\end{figure}

In the next section, results (populations of polaritonic states and photon emission signal)
with several different cavity and laser parameters will be presented.
In all cases, the system is prepared in a pure initial state with
$\hat{\rho}(t=0) = |\Psi_0 \rangle \langle \Psi_0|$ (Lindblad) or
$|\Psi(t=0) \rangle = |\Psi_0 \rangle$ (TDSE) where $|\Psi_0 \rangle$ is the ground state of
the coupled cavity-molecule system. Then, the system is excited by a pump laser pulse
which in most cases is chosen such that it transfers population primarily to the
singly-excited subspace, either to the 1LP or to the 1UP states.
One-dimensional diabatic PES cuts along the vibrational mode $\nu_2$
and two-dimensional 1LP and 1UP PESs are depicted in
Fig. \ref{fig:PES} for the cavity parameters
$\omega_{\textrm{c}}=29957.2~\textrm{cm}^{-1}$ and $g=0.02~\textrm{au}$.
Occasionally, higher-lying polaritonic states will be populated as well, which will shed
further light on the difference between the Lindblad and TDSE-based descriptions.

\section{Results and discussion}
\label{sec:results}

Having described the theory, the computational protocol and the molecular system under
investigation, we will now examine the populations of polaritonic states as well as
the ultrafast radiative emission signal of the cavity. We will mainly focus on the 1LP and
1UP polaritonic states of the singly-excited subspace. We will discuss in detail how
the different pump laser parameters (frequency, intensity, and pulse length) as well as
the frequency, coupling strength, and decay rate of the cavity mode influence the physical
quantities of interest.

\subsection{Population transfer to the lower polariton (1LP state)}
\label{sec:results_case1}

\begin{figure}
\includegraphics[width=0.495\textwidth]{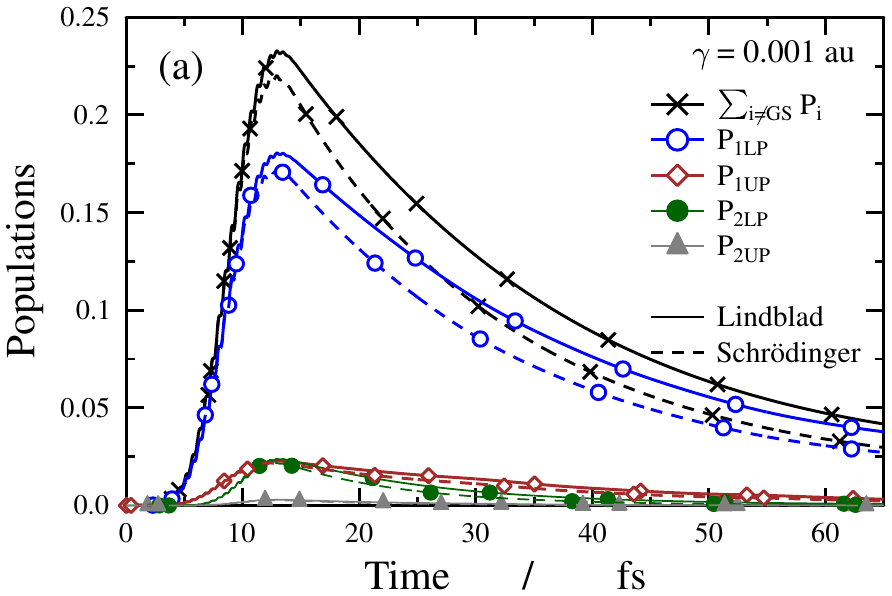}
\includegraphics[width=0.495\textwidth]{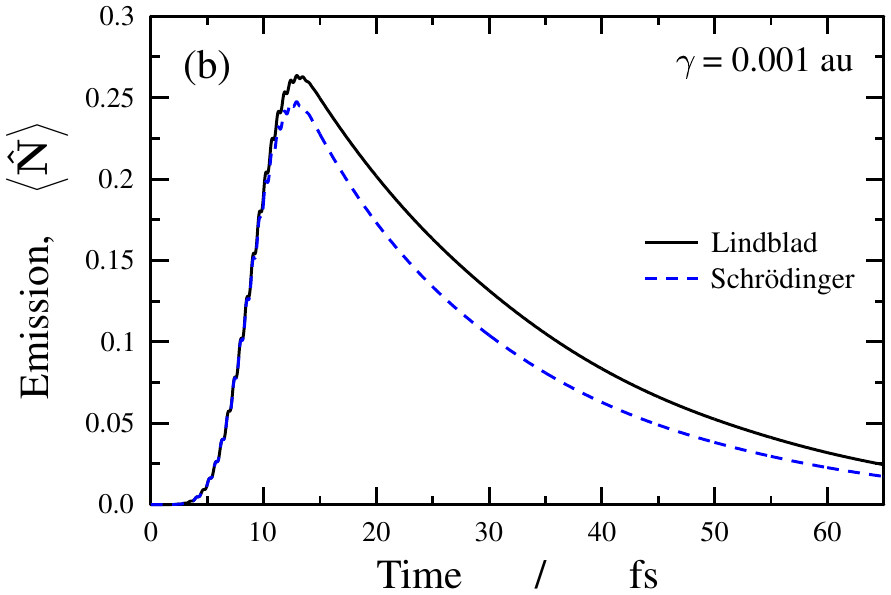}
\includegraphics[width=0.495\textwidth]{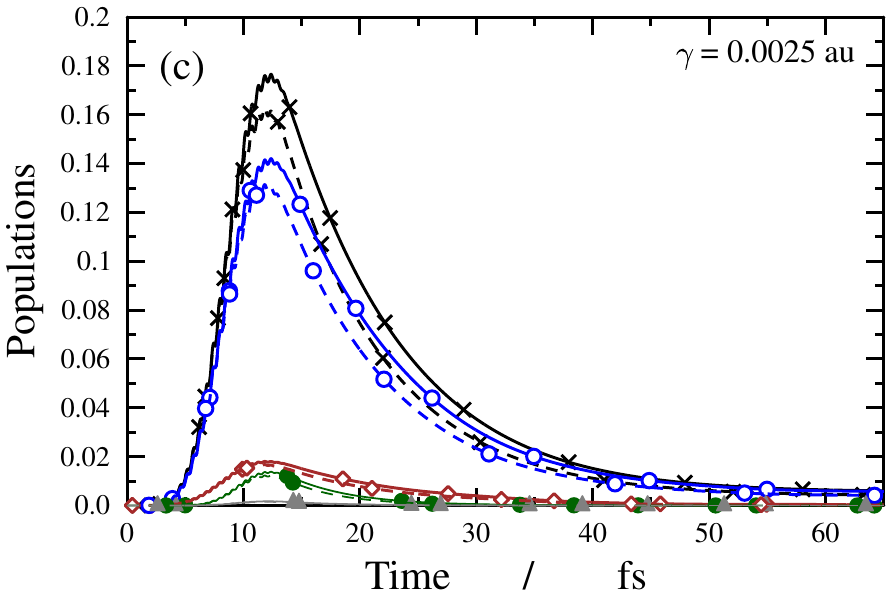}
\includegraphics[width=0.495\textwidth]{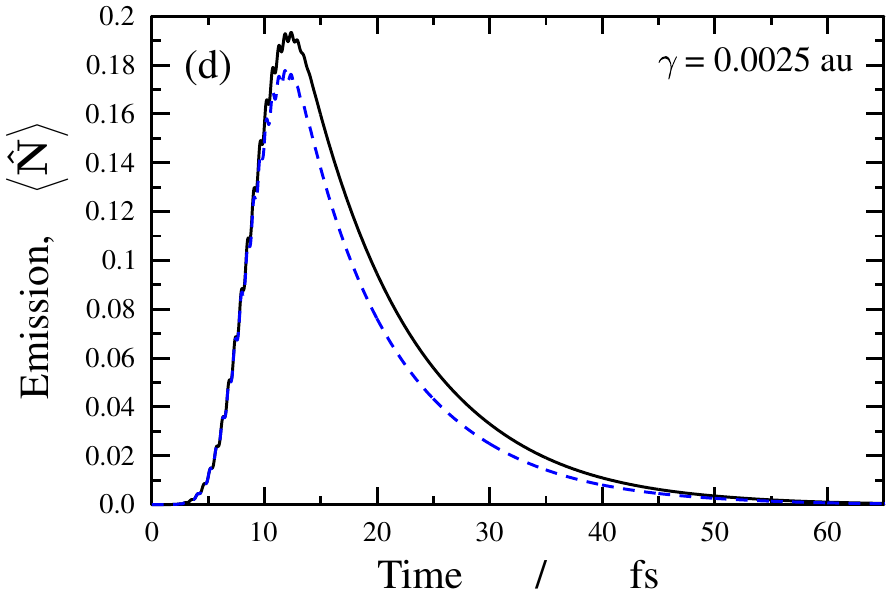}
\includegraphics[width=0.495\textwidth]{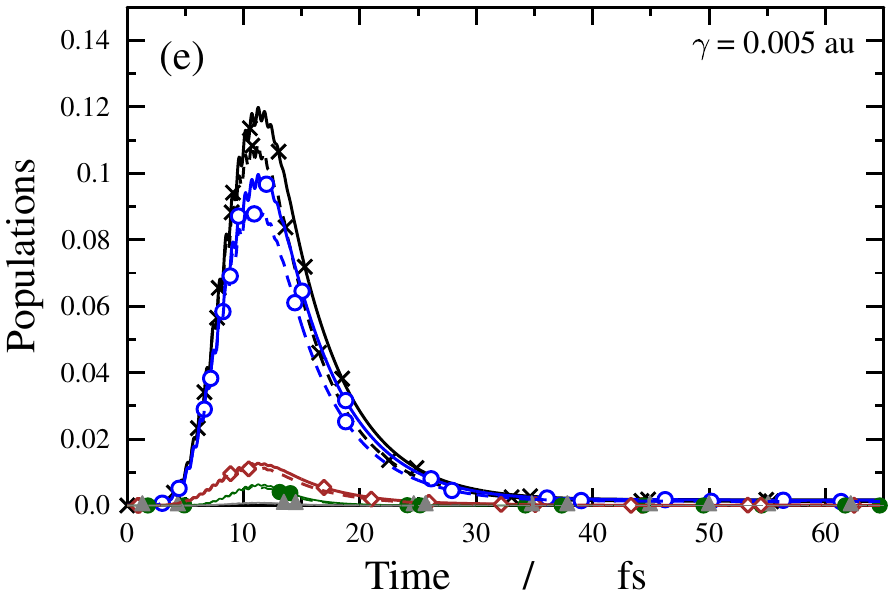}
\includegraphics[width=0.495\textwidth]{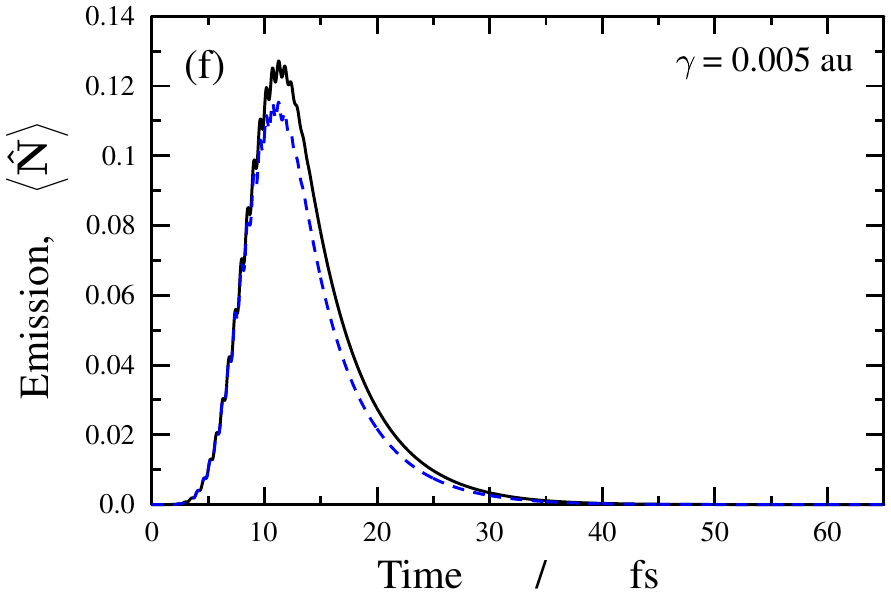}
\caption{\label{fig:LP_popem_1} 
Populations of polaritonic states and emission curves
(the emission is proportional to the expectation value of the photon number operator $\hat{N}$)
for cavity decay rates $\gamma_{\textrm{c}}=0.001~\textrm{au}$
(equivalent to a lifetime of $\tau = 24.2~\textrm{fs}$, panels a and b), 
$\gamma_{\textrm{c}}=0.0025~\textrm{au}$
($\tau = 9.7~\textrm{fs}$, panels c and d) and 
$\gamma_{\textrm{c}}=0.005~\textrm{au}$
($\tau = 4.8~\textrm{fs}$, panels e and f). 
The cavity wavenumber and coupling strength
equal $\omega_{\textrm{c}}=29957.2~\textrm{cm}^{-1}$
and $g=0.01~\textrm{au}$, respectively. Parameters of the pump
laser pulse are chosen as $\omega=30000~\textrm{cm}^{-1}$,
$T=15~\textrm{fs}$ and $I=5\cdot10^{11}~\textrm{W}/\textrm{cm}^{2}$.
The Lindblad results (solid lines) agree well with their
Schr\"odinger (TDSE) counterparts (dashed lines) in all cases presented.}
\end{figure}

\begin{figure}
\includegraphics[width=0.495\textwidth]{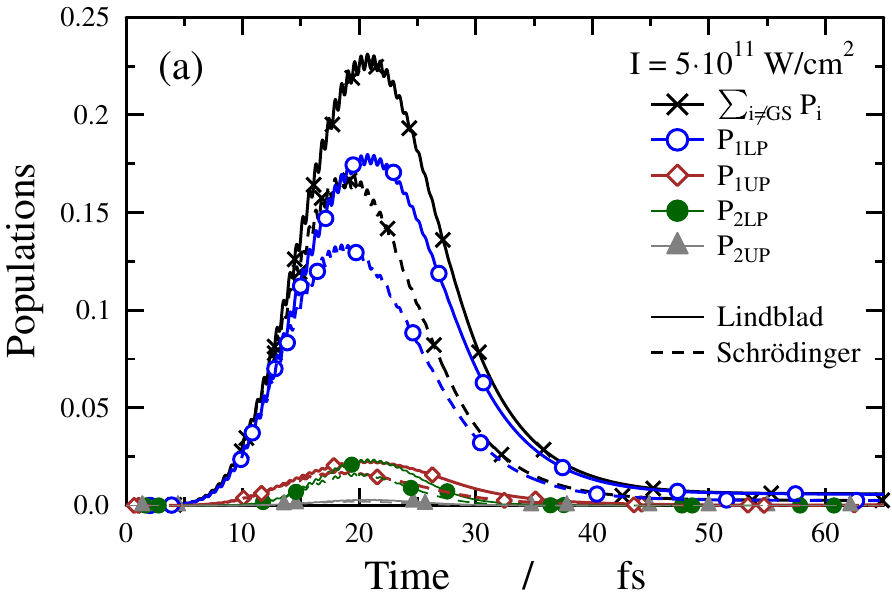}
\includegraphics[width=0.495\textwidth]{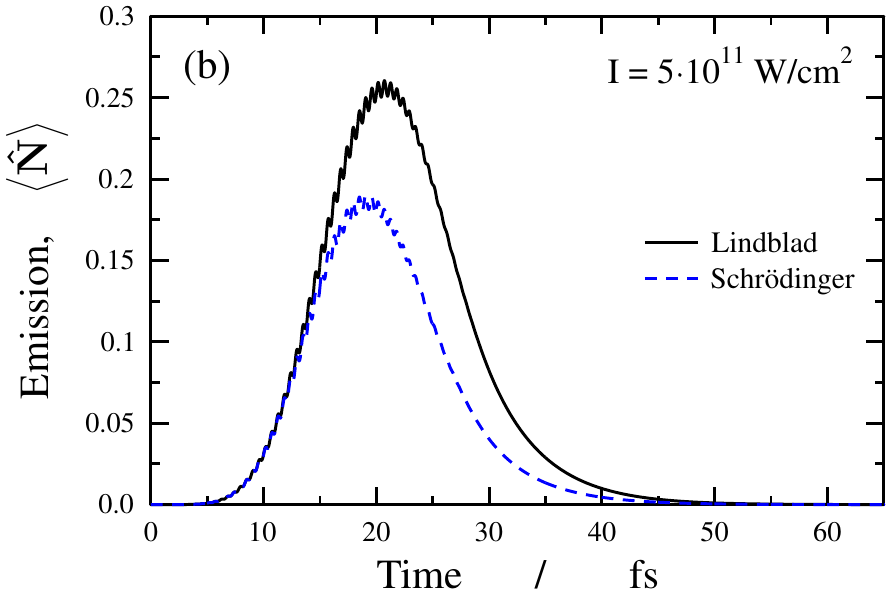}
\includegraphics[width=0.495\textwidth]{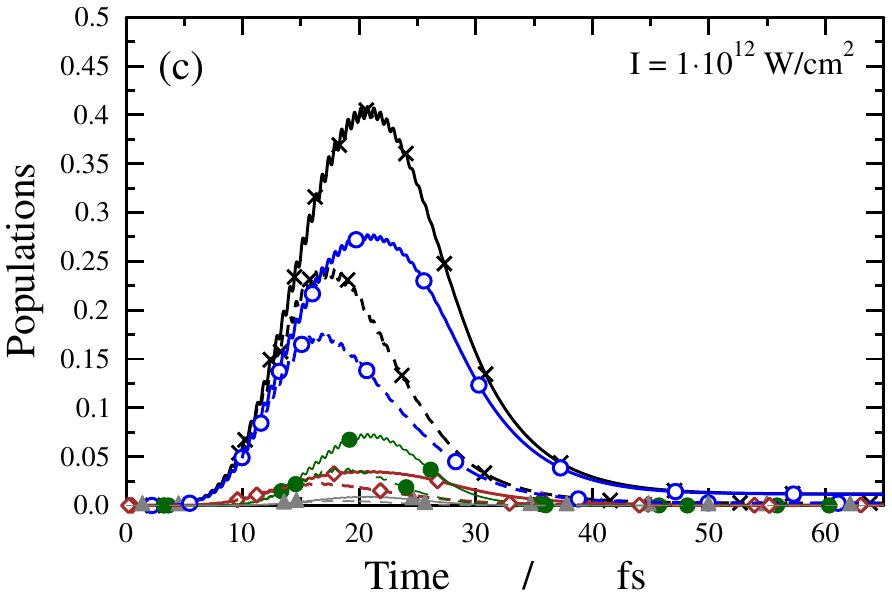}
\includegraphics[width=0.495\textwidth]{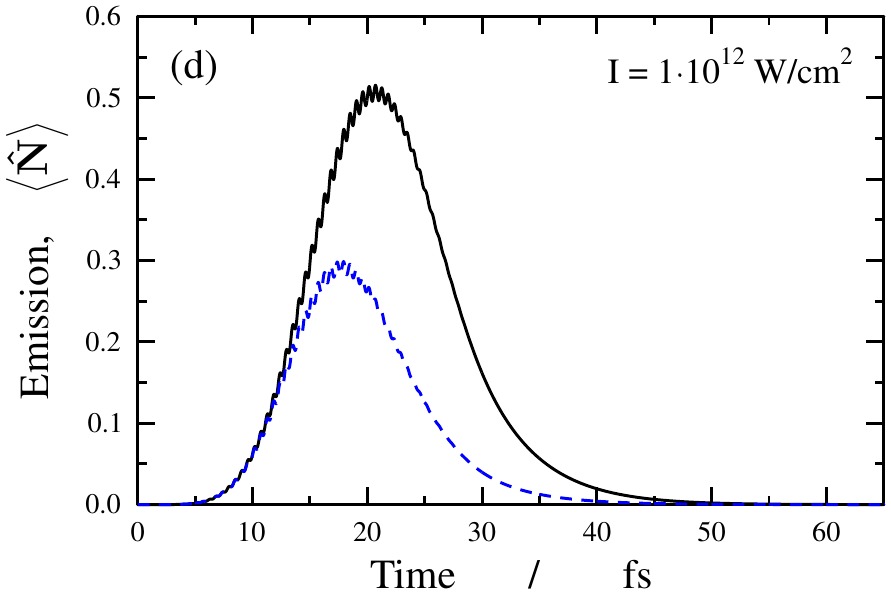}
\caption{\label{fig:LP_popem_2} 
Populations of polaritonic states and emission curves
(the emission is proportional to the expectation value of the photon number operator $\hat{N}$)
for laser intensities $I=5\cdot10^{11}~\textrm{W}/\textrm{cm}^{2}$
(panels a and b) and $I=10^{12}~\textrm{W}/\textrm{cm}^{2}$
(panels c and d). The cavity wavenumber and coupling strength
equal $\omega_{\textrm{c}}=29957.2~\textrm{cm}^{-1}$
and $g=0.01~\textrm{au}$, respectively, while the cavity decay
rate is set to $\gamma_{\textrm{c}}=0.005~\textrm{au}$
(equivalent to a lifetime of $\tau = 4.8~\textrm{fs}$).
Other parameters of the pump laser pulse are chosen as
$\omega=30000~\textrm{cm}^{-1}$ and $T=30~\textrm{fs}$.
The Lindblad results (solid lines) show visible deviations from
their Schr\"odinger (TDSE) counterparts (dashed lines).}
\end{figure}

\begin{figure}
\includegraphics[width=0.495\textwidth]{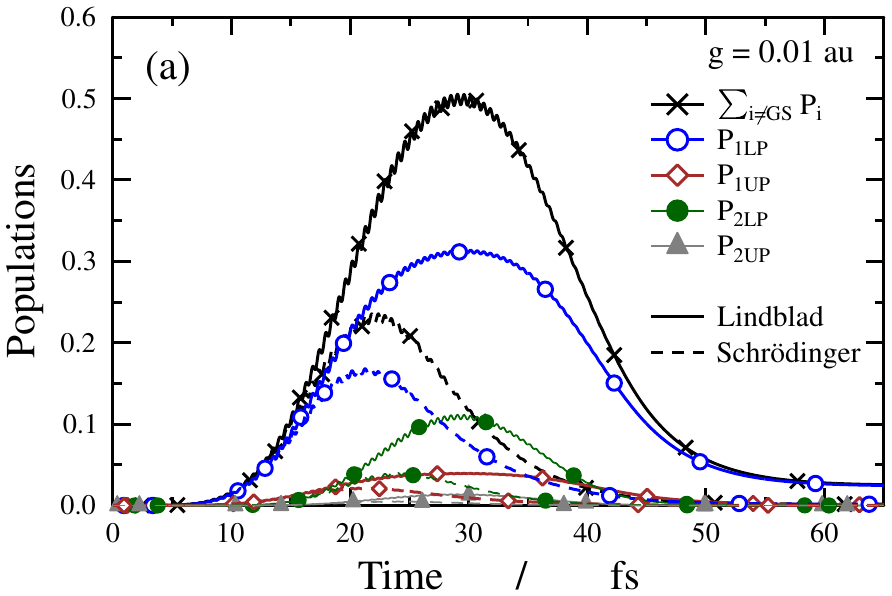}
\includegraphics[width=0.495\textwidth]{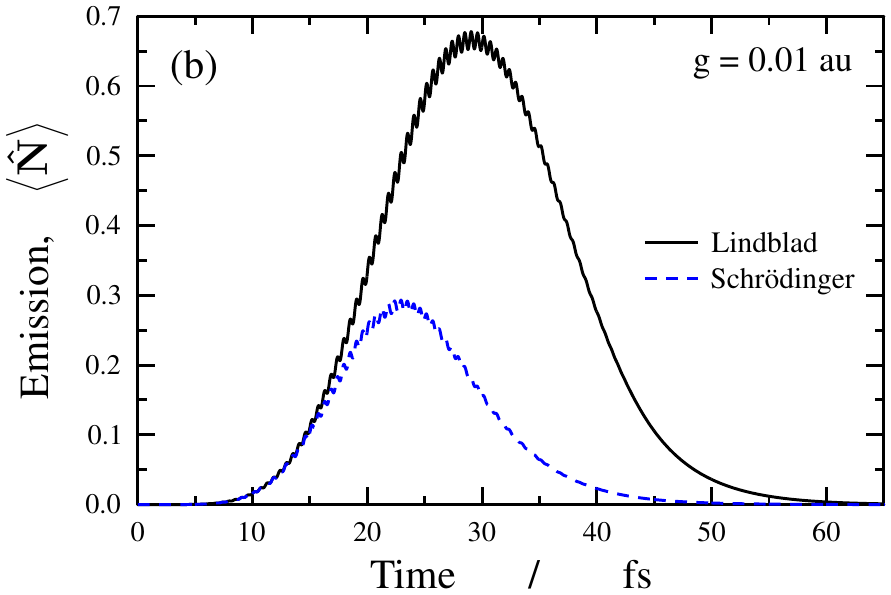}
\includegraphics[width=0.495\textwidth]{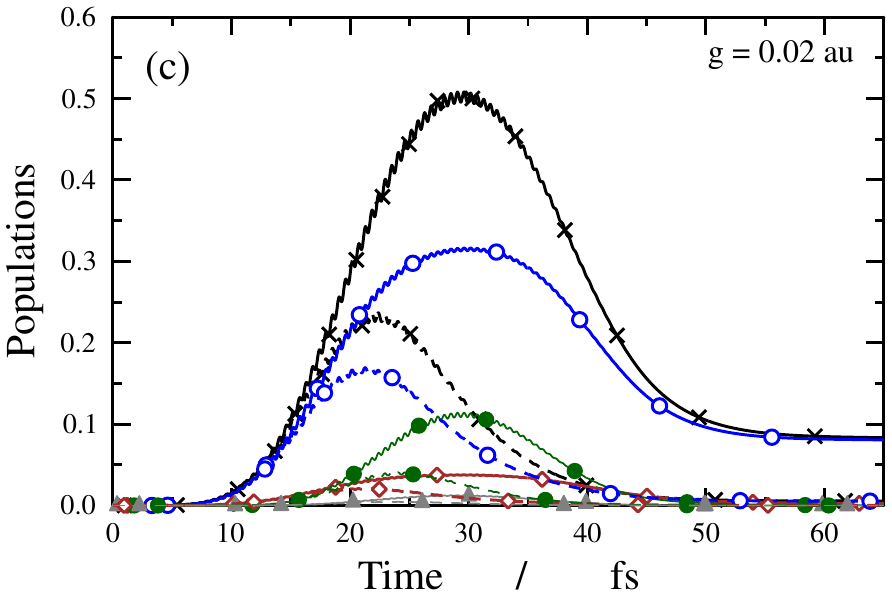}
\includegraphics[width=0.495\textwidth]{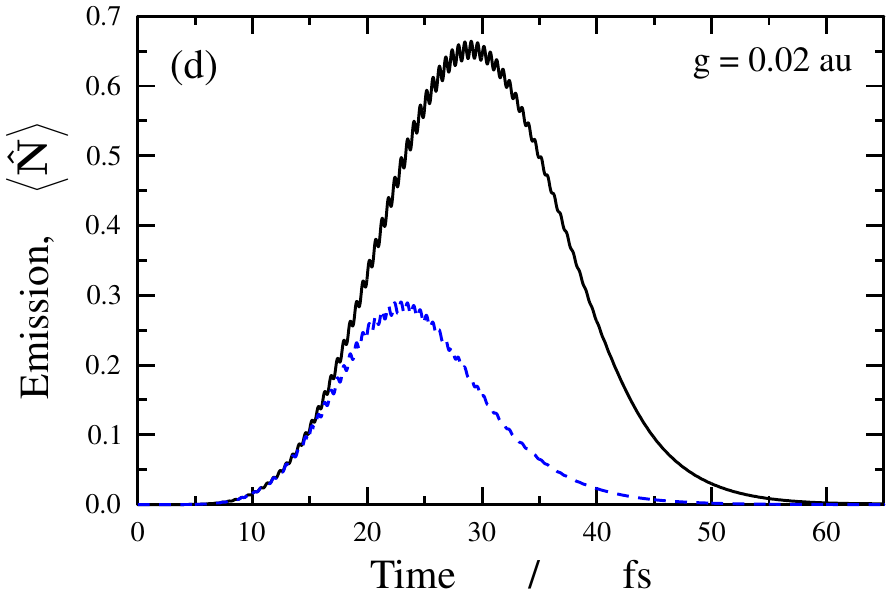}
\caption{\label{fig:LP_popem_3}
Populations of polaritonic states and emission curves (the emission is proportional to the
expectation value of the photon number operator $\hat{N}$) 
for coupling strength values of
$g=0.01~\textrm{au}$ (panels a and b) and $g=0.02~\textrm{au}$ (panels c and d).
The cavity wavenumber and decay rate equal $\omega_{\textrm{c}}=29957.2~\textrm{cm}^{-1}$
and $\gamma_{\textrm{c}}=0.005~\textrm{au}$
(equivalent to a lifetime of $\tau = 4.8~\textrm{fs}$),
respectively.
Parameters of the pump laser pulse are chosen as $\omega=30000~\textrm{cm}^{-1}$,
$T=45~\textrm{fs}$ and  $I=10^{12}~\textrm{W}/\textrm{cm}^{2}$.
The Lindblad results (solid lines) differ substantially from their Schr\"odinger (TDSE)
counterparts (dashed lines).}
\end{figure}

\begin{figure}
\includegraphics[width=0.495\textwidth]{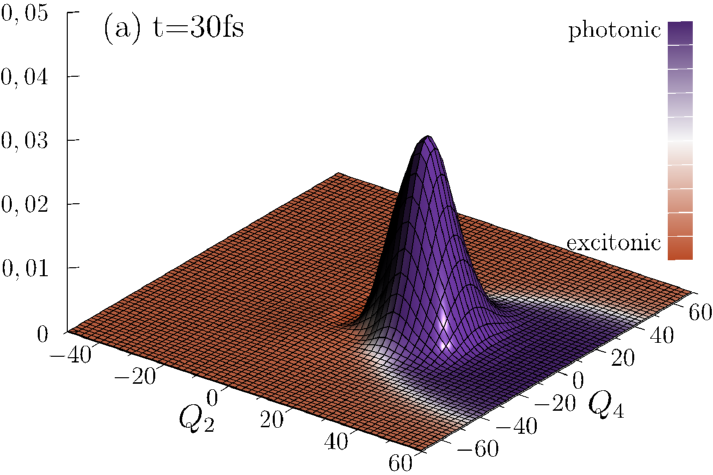}
\includegraphics[width=0.495\textwidth]{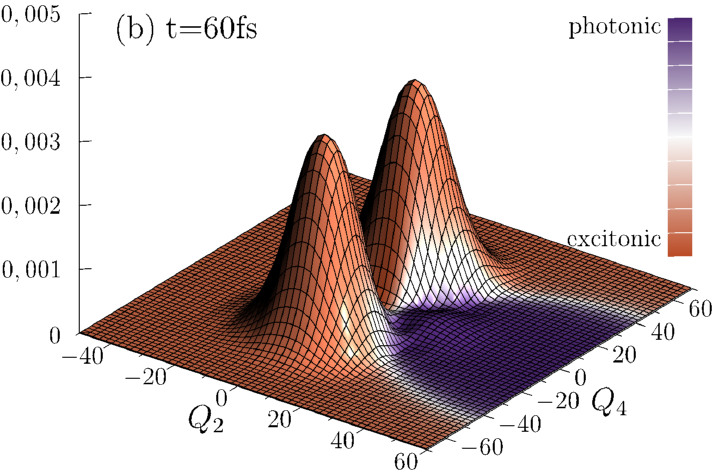}
\caption{\label{fig:LP_dens}
Probability density figures (obtained with the Lindblad equation)
for the lower polaritonic (1LP) potential energy surface 
at $t=30~\textrm{fs}$ (panel a) and $t=60~\textrm{fs}$ (panel b). 
The cavity wavenumber, coupling strength and decay rate equal
$\omega_{\textrm{c}}=29957.2~\textrm{cm}^{-1}$, $g=0.02~\textrm{au}$
and $\gamma_{\textrm{c}}=0.005~\textrm{au}$, respectively.
Parameters of the pump laser pulse are chosen as
$\omega=30000~\textrm{cm}^{-1}$, $T=45~\textrm{fs}$ and 
$I=10^{12}~\textrm{W}/\textrm{cm}^{2}$.
The laser pulse mainly populates the 1LP state (and also the
2LP state to a lesser extent) in this case.
At $t=30~\textrm{fs}$ and $t=60~\textrm{fs}$ the wave packet is
localized in the photonic (purple) and excitonic (orange) regions of the 1LP PES,
respectively.}
\end{figure}

In this section all numerical results (from Fig. \ref{fig:LP_popem_1} to Fig. \ref{fig:LP_dens})
have been obtained with $\omega=30000~\textrm{cm}^{-1}$ (carrier wavenumber of the pump laser pulse)
and $\omega_{\textrm{c}}=29957.2~\textrm{cm}^{-1}$ (cavity wavenumber), while the remaining laser
and cavity parameters vary from figure to figure. We note that the current $\omega$ and 
$\omega_\textrm{c}$ values correspond to case 1 of Ref. \onlinecite{22FaHaVi} where oscillations
in the emission signal were attributed to wave packet motion between the photonic
(ground-state molecule with one photon) and excitonic (excited-state molecule with zero photons)
parts of the 1LP PES.

As for the laser parameters in Fig. \ref{fig:LP_popem_1}, we have applied the values
$T=15~\textrm{fs}$ (pulse length) and $I=5\cdot10^{11}~\textrm{W}/\textrm{cm}^{2}$ (intensity). Regarding the cavity coupling strength and decay rate parameters, the values $g=0.01~\textrm{au}$
as well as $\gamma_{\textrm{c}}=0.001~\textrm{au}$,
$\gamma_{\textrm{c}}=0.0025~\textrm{au}$ and $\gamma_{\textrm{c}}=0.005~\textrm{au}$ are used, respectively. The latter values are equivalent to lifetimes of 
$\tau = 1/\gamma_{\textrm{c}} = 24.2~\textrm{fs}$ ($\gamma_{\textrm{c}}=0.001~\textrm{au}$), 
$\tau = 9.7~\textrm{fs}$ ($\gamma_{\textrm{c}}=0.0025~\textrm{au}$) and
$\tau = 4.8~\textrm{fs}$ ($\gamma_{\textrm{c}}=0.005~\textrm{au}$).
In this case, the laser transfers population to the photonic part of the 1LP state with high
selectivity and the ensuing quantum dynamics takes place almost entirely on the 1LP PES.
This is confirmed by Fig. \ref{fig:LP_popem_1} which shows 
that polaritonic states above the 1LP state are hardly populated.
In Fig. \ref{fig:LP_popem_1} the labels 2LP and 2UP denote polaritonic states that
can be characterized as superpositions of the diabatic states $|\textrm{X}2\rangle$ and 
$|\textrm{A}1\rangle$ to a good approximation.
The amount of population transferred to the 1LP state depends on the lifetime
of the cavity photon and also on the method (Lindblad or TDSE) used.
In case of the Lindblad approach, 
due to the incoherent transition $|\textrm{X}1\rangle \rightarrow |\textrm{X}0\rangle$ induced
by the term $\gamma_\textrm{c} \hat{a} \hat{\rho} \hat{a}^\dag$, part of the population is
transferred back to the ground polaritonic state from 1LP already during the pumping process.
This is not possible with the non-Hermitian TDSE, as the 1LP population absorbed by
the imaginary part of the non-Hermitian Hamiltonian is not restored to the ground polaritonic
state. Not surprisingly, higher values of $\gamma_\textrm{c}$ lead to faster relaxation
of the excited polaritonic populations and the emission signal.
Larger values of $\gamma_\textrm{c}$ are generally expected to result in larger
discrepancies between the Lindblad and TDSE-based descriptions.
This tendency is a direct consequence that the incoherent decay term is missing in 
the non-Hermitian TDSE scheme. At the same time, however, the 1LP populations and corresponding
emissions agree well for the parameter ranges investigated in Fig. \ref{fig:LP_popem_1}.

Fig. \ref{fig:LP_popem_2} presents results obtained with parameters different
from the ones used in Fig. \ref{fig:LP_popem_1}. Here we have employed the highest cavity
decay rate of $\gamma_{\textrm{c}}=0.005~\textrm{au}$, $g=0.01~\textrm{au}$, laser intensities
of $I=5\cdot10^{11}~\textrm{W}/\textrm{cm}^{2}$ and $I=10^{12}~\textrm{W}/\textrm{cm}^{2}$
and doubled the pulse duration to $T=30~\textrm{fs}$ from $T=15~\textrm{fs}$.
The results obtained differ significantly from those displayed in Fig. \ref{fig:LP_popem_1}.
As the laser pulse gets longer and its intensity gets higher, the non-Hermitian TDSE method
breaks down spectacularly even for the populations of excited polaritonic states and for the
emission signal. The reason for these observations is that the non-Hermitian TDSE
does not return population to the ground polaritonic state by the incoherent decay term.
As already stated, this is not the case for the Lindblad master equation.  If the length of
the pulse is sufficiently large compared to the cavity lifetime, the incoherent decay term 
starts to repopulate the ground polaritonic state before the laser pulse ends.
This way, the population returned to the ground polaritonic state can be transferred again to
excited polaritonic states (dominantly to the 1LP state in this case), which results in a 
more efficient population transfer compared to the TDSE approach.
This finding explains why the Lindblad 1LP population and emission values exceed their
TDSE counterparts in Fig. \ref{fig:LP_popem_2}.
One can also notice that there is some population transfer above the
singly-excited subspace. This is potentially interesting as the decay is faster
for larger photon numbers and in the Lindblad case population in the singly-excited subspace
can be replenished from higher-lying polaritonic states by the incoherent decay term.

The trends observed in Fig. \ref{fig:LP_popem_2} are even more pronounced in 
Fig. \ref{fig:LP_popem_3}. Here the pump pulse is even longer ($T=45~\textrm{fs}$) with 
$I=10^{12}~\textrm{W}/\textrm{cm}^{2}$ and the coupling strength is doubled from
$g=0.01~\textrm{au}$ to $g=0.02~\textrm{au}$ with $\gamma_{\textrm{c}}=0.005~\textrm{au}$.
We can again observe the breakdown of the non-Hermitian TDSE method and
there is visible population transfer above the singly-excited subspace.
Moreover, panel c of Fig. \ref{fig:LP_popem_3} shows another interesting effect,
namely, for $g=0.02~\textrm{au}$ a certain amount of the population
is trapped in the 1LP PES in the Lindblad case. Such trapping is also noticeable in panel a
of Fig. \ref{fig:LP_popem_3} ($g=0.01~\textrm{au}$), although to a lesser extent.
The use of longer pulses and higher laser intensities facilitates the population
of higher excited polaritonic states (mainly the 2LP state), from where the Lindblad scheme
can transfer population back to the singly-excited subspace.
In addition, the laser pulse can also transfer some population already relaxed to the ground
polaritonic state back to 1LP since the pulse length $T$ is considerably larger than
the cavity lifetime $\tau$ ($T = 45~\textrm{fs} > \tau = 4.8~\textrm{fs}$ in this case).
The combination of these two effects gives rise to appreciable Lindblad 1LP population
at the end of the laser pulse, while in the non-Hermitian TDSE case the 1LP is essentially
zero at $t=45~\textrm{fs}$. The origin of the trapping mechanism is elucidated by
probability density figures shown in Fig. \ref{fig:LP_dens}. The Lindblad probability
densities reveal that the pump pulse indeed transfers population to the photonic region of the
1LP PES (see panel a of Fig. \ref{fig:LP_dens} at $t=30~\textrm{fs}$).
Subsequently, the wave packet splits into two parts and moves to the excitonic region of the
1LP PES (panel b, $t=60~\textrm{fs}$), which quenches photon emission and results in
the trapping effect mentioned. Of course, for longer times, the wave packet will
move back to the photonic region, leading to decay to the ground polaritonic state.
It is conspicuous in Fig. \ref{fig:PES} (panel b) that the 1LP PES has two symmetry-equivalent
local minima which can trap the wave packet in the excitonic region for some time.

We have repeated the TDSE calculations of this section by setting norm of the
wave function to one at every time step (renormalized TDSE approach).  Results are presented
in Appendix \ref{sec:appendixA}  (for population and emission results see Figs.
\ref{fig:LP_popem_1_renorm},  \ref{fig:LP_popem_2_renorm} and \ref{fig:LP_popem_3_renorm} 
which correspond to the parameters of Figs. \ref{fig:LP_popem_1},  \ref{fig:LP_popem_2} and
\ref{fig:LP_popem_3}, respectively). It is clearly visible that the renormalized TDSE method
greatly reduces the discrepancy between the Lindblad and TDSE results. This finding can be 
attributed to the renormalization of the wave function, which refills the population of the
ground (lowest) polaritonic state. In addition, the purity of the density matrix, 
$\textrm{tr}(\hat{\rho}^2)$, is also given as a function of time in Appendix \ref{sec:appendixA}
(see Figs. \ref{fig:LP_purity_1}, \ref{fig:LP_purity_2} and \ref{fig:LP_purity_3}).
In all cases considered in this section, the purity is exactly one at $t=0$ and remains
close to one during and after excitation.

Finally, reference results with a special initial state (molecule in the
vibrational ground state of the ground (X) electronic state and 1 photon in the cavity mode)
are outlined (see Appendix \ref{sec:appendixB}). The cavity parameters are chosen as 
$\omega_{\textrm{c}}=29957.2~\textrm{cm}^{-1}$, $g=0.01~\textrm{au}$ and
$\gamma_{\textrm{c}}=0.001~\textrm{au}$. With this special setup, the singly-excited
subspace (1LP and 1UP states) is populated at $t=0$. Fig. \ref{fig:special_1} compares 
populations of polaritonic states (panel a) and emission curves (panel b) obtained with the
Lindblad and non-Hermitian TDSE methods, showing good agreement. On the contrary, the
Lindblad and renormalized TDSE populations and emissions (see panels a and b of Fig.
\ref{fig:special_2}) show substantial deviations. The breakdown of the renormalized TDSE method 
can be readily rationalized by keeping in mind that the ground polaritonic state initially bears
negligible population. In this case, renormalization of the wave function will keep the 
population trapped in the singly-excited subspace and the ground polaritonic state can not
be refilled. This is to be contrasted with the previous cases where the laser pulse has 
transferred a certain amount of population from the ground polaritonic state to the 
singly-excited subspace and occasionally to higher-lying polaritonic states.
Finally, Fig. \ref{fig:special_purity} presents purity results (see the curve labeled as LP).
As the initial state of the system is a pure state, the purity equals one at $t=0$. Later,
the system evolves into a mixed state and the purity gradually decreases to about $0.5$.
As relaxation continues to transfer population to the ground polaritonic state, the purity
starts increasing again.

\subsection{Population transfer to the upper polariton (1UP state)}
\label{sec:results_case2}

\begin{figure}
\includegraphics[width=0.495\textwidth]{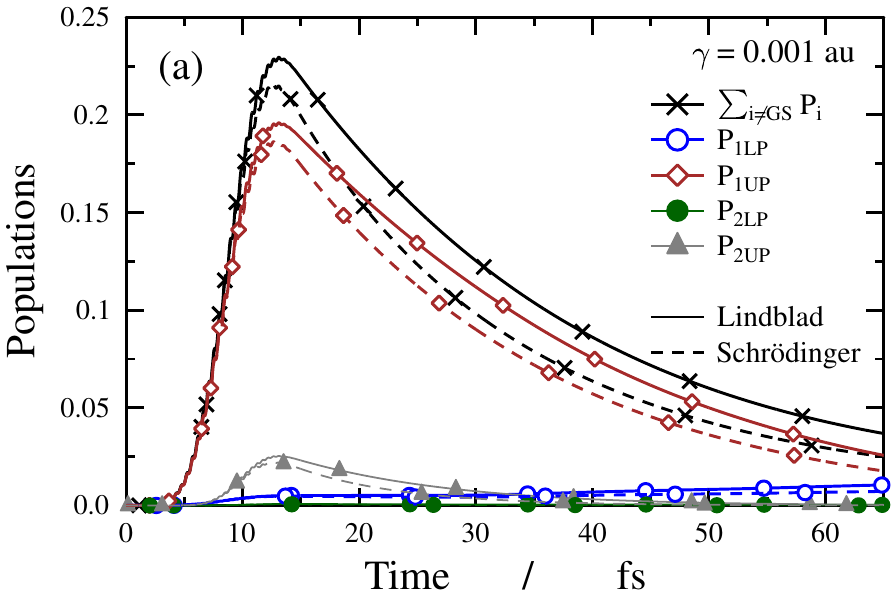}
\includegraphics[width=0.495\textwidth]{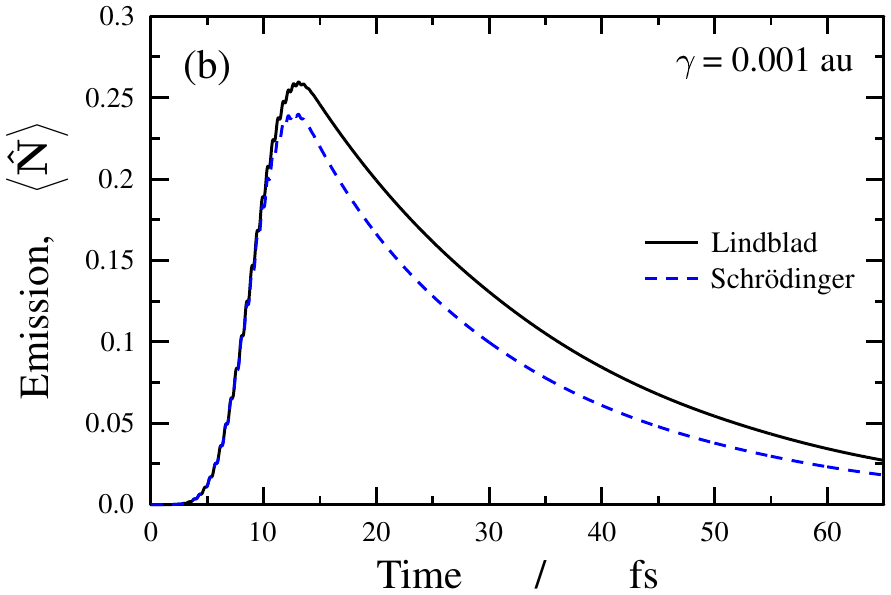}
\includegraphics[width=0.495\textwidth]{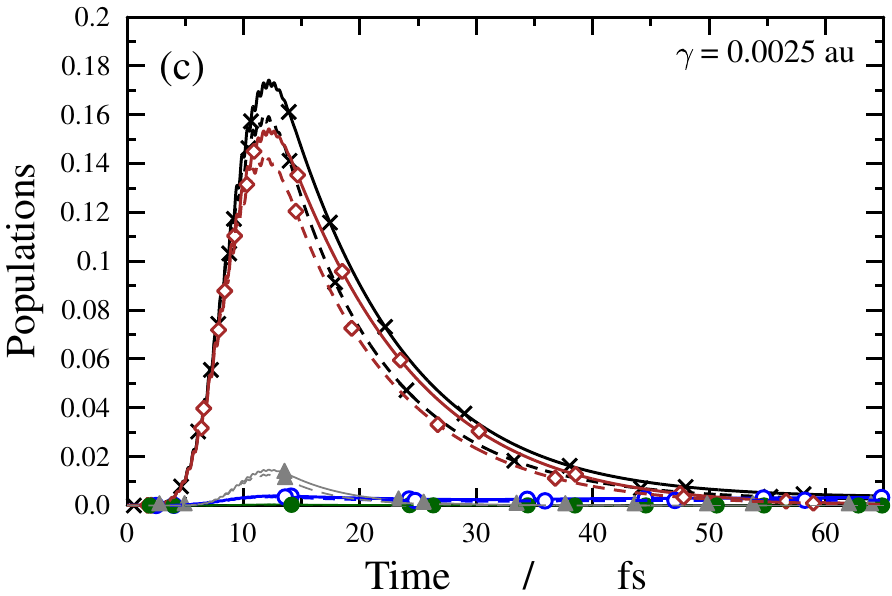}
\includegraphics[width=0.495\textwidth]{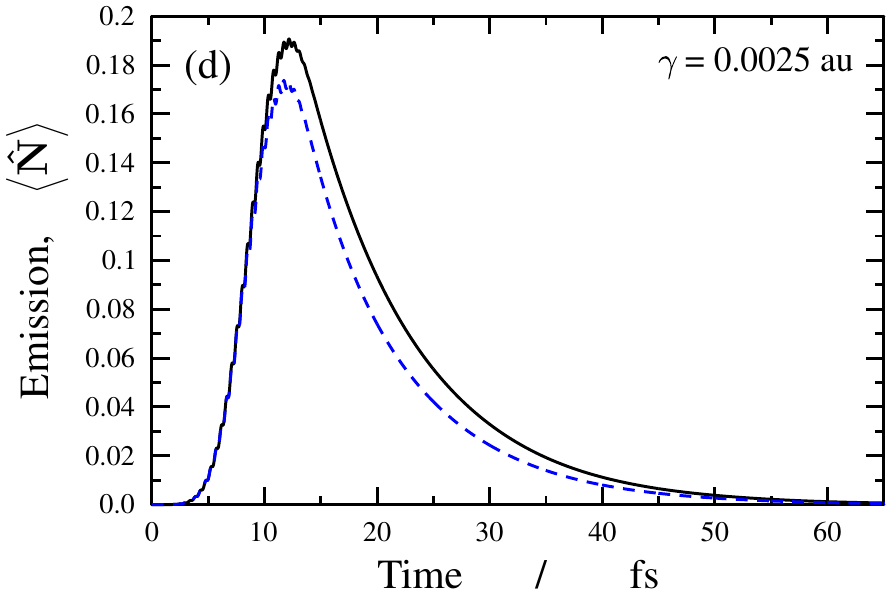}
\includegraphics[width=0.495\textwidth]{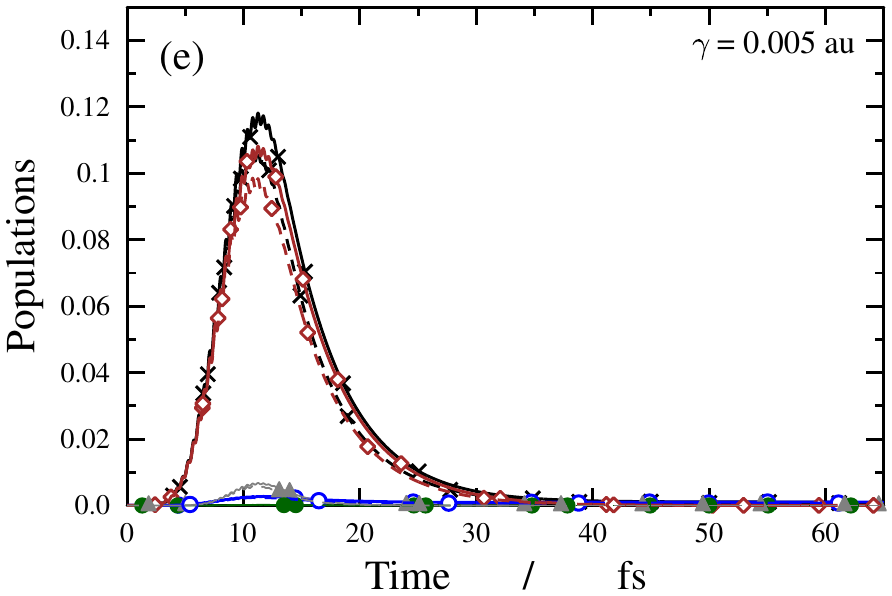}
\includegraphics[width=0.495\textwidth]{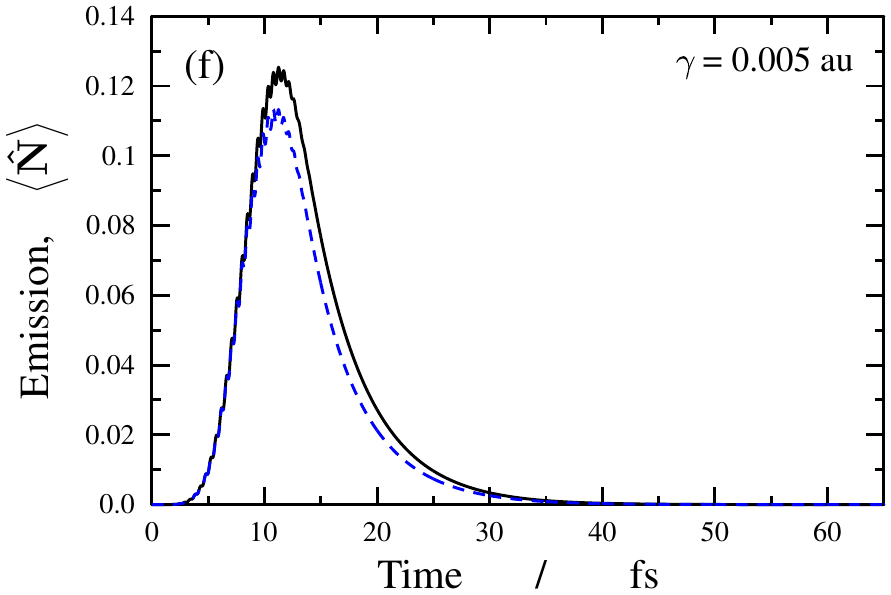}
\caption{\label{fig:UP_popem_1}
Populations of polaritonic states and emission curves
(the emission is proportional to the expectation value of the photon number operator $\hat{N}$)
for cavity decay rates $\gamma_{\textrm{c}}=0.001~\textrm{au}$
(equivalent to a lifetime of $\tau = 24.2~\textrm{fs}$,
panels a and b), 
$\gamma_{\textrm{c}}=0.0025~\textrm{au}$
($\tau = 9.7~\textrm{fs}$, panels c and d) and 
$\gamma_{\textrm{c}}=0.005~\textrm{au}$
($\tau = 4.8~\textrm{fs}$, panels e and f). 
The cavity wavenumber and coupling strength
equal $\omega_{\textrm{c}}=35744.8~\textrm{cm}^{-1}$
and $g=0.01~\textrm{au}$, respectively. Parameters of the pump
laser pulse are chosen as $\omega=36000~\textrm{cm}^{-1}$,
$T=15~\textrm{fs}$ and $I=5\cdot10^{11}~\textrm{W}/\textrm{cm}^{2}$.
The Lindblad results (solid lines) agree well with their
Schr\"odinger (TDSE) counterparts (dashed lines) in all cases presented.}
\end{figure}

\begin{figure}
\includegraphics[width=0.495\textwidth]{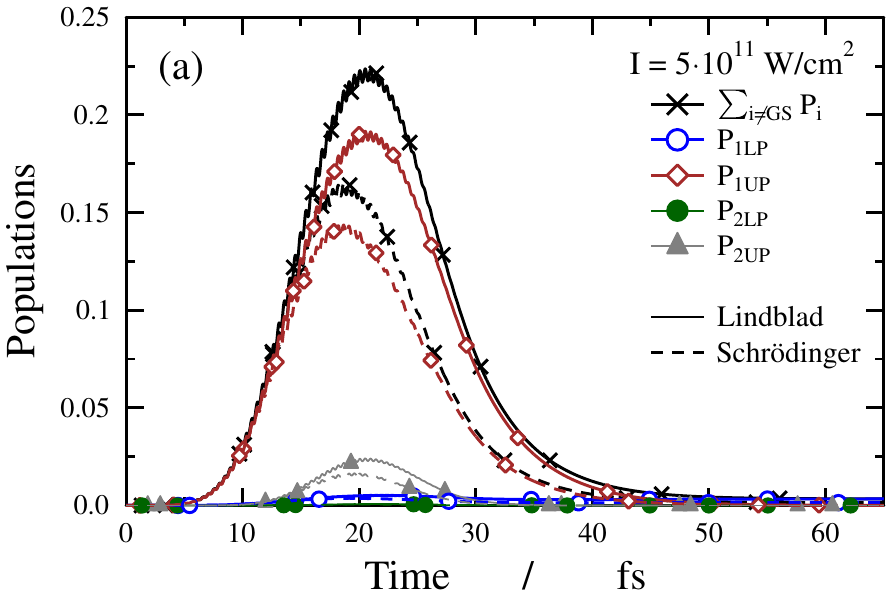}
\includegraphics[width=0.495\textwidth]{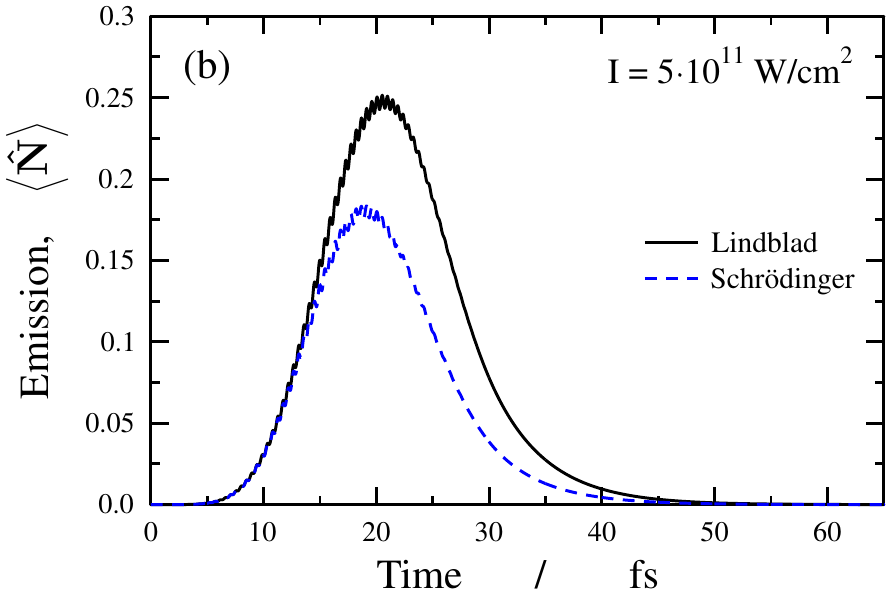}
\includegraphics[width=0.495\textwidth]{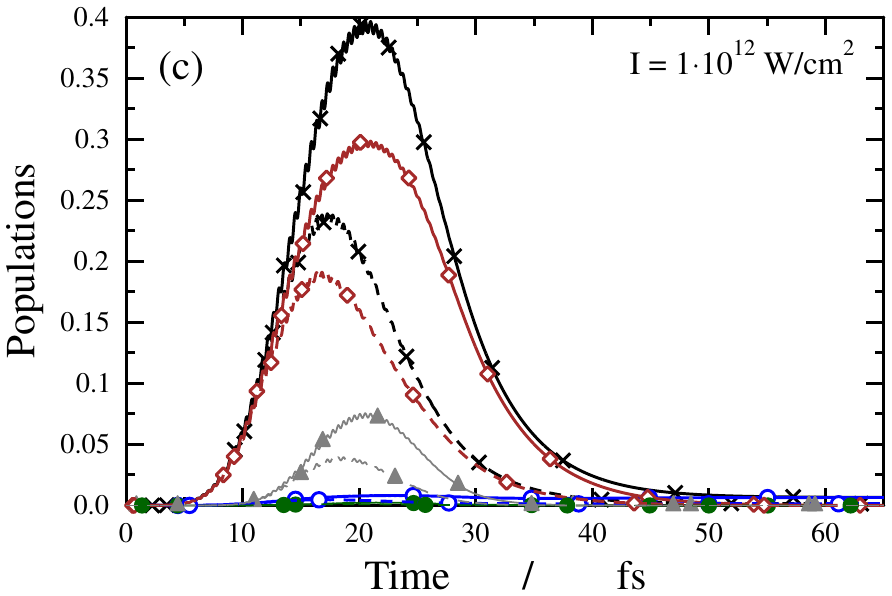}
\includegraphics[width=0.495\textwidth]{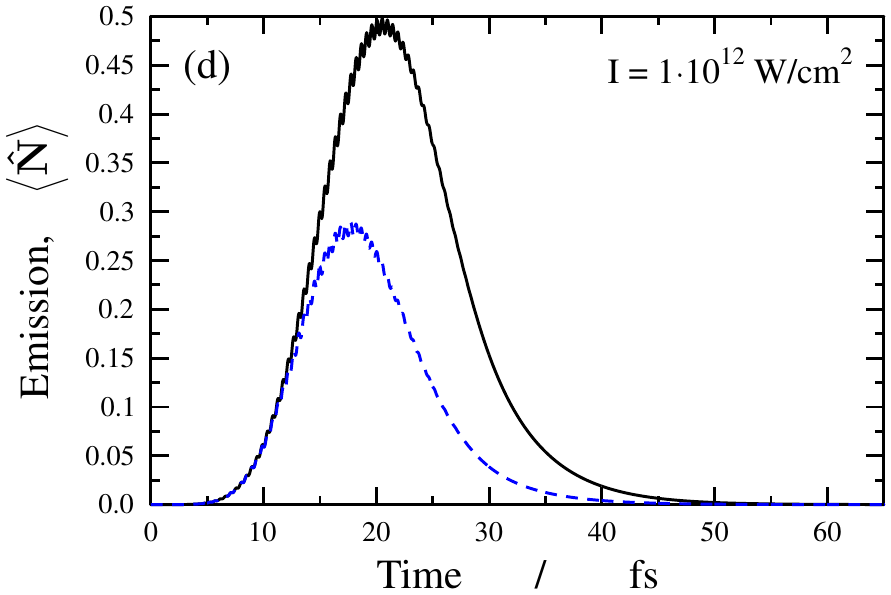}
\caption{\label{fig:UP_popem_2}
Populations of polaritonic states and emission curves 
(the emission is proportional to the expectation value of the photon number operator $\hat{N}$)
for laser intensities $I=5\cdot10^{11}~\textrm{W}/\textrm{cm}^{2}$
(panels a and b) and $I=10^{12}~\textrm{W}/\textrm{cm}^{2}$
(panels c and d). The cavity wavenumber and coupling strength
equal $\omega_{\textrm{c}}=35744.8~\textrm{cm}^{-1}$
and $g=0.01~\textrm{au}$, respectively, while the cavity decay
rate is set to $\gamma_{\textrm{c}}=0.005~\textrm{au}$
(equivalent to a lifetime of $\tau = 4.8~\textrm{fs}$).
Other parameters of the pump laser pulse are chosen as
$\omega=36000~\textrm{cm}^{-1}$ and $T=30~\textrm{fs}$.
The Lindblad results (solid lines) show visible deviations from
their Schr\"odinger (TDSE) counterparts (dashed lines).}
\end{figure}

\begin{figure}
\includegraphics[width=0.495\textwidth]{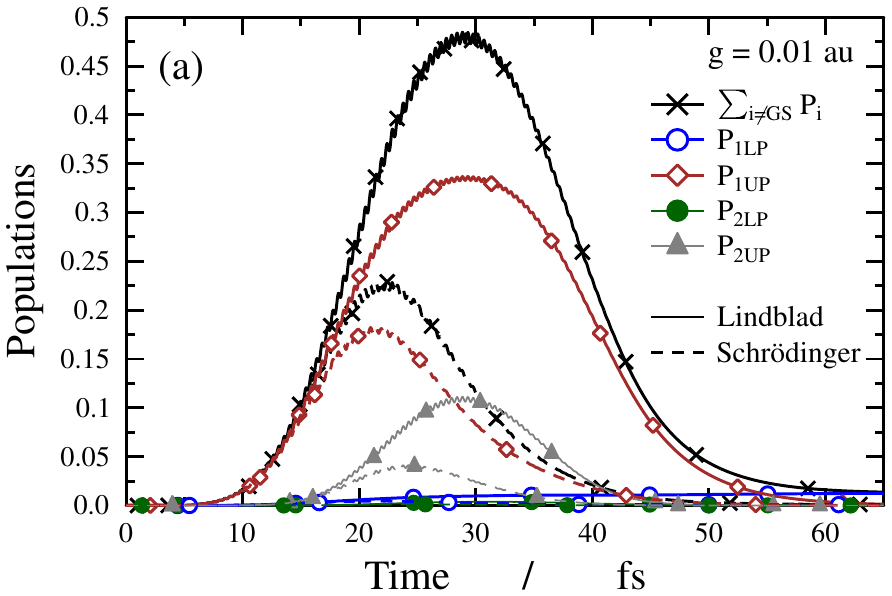}
\includegraphics[width=0.495\textwidth]{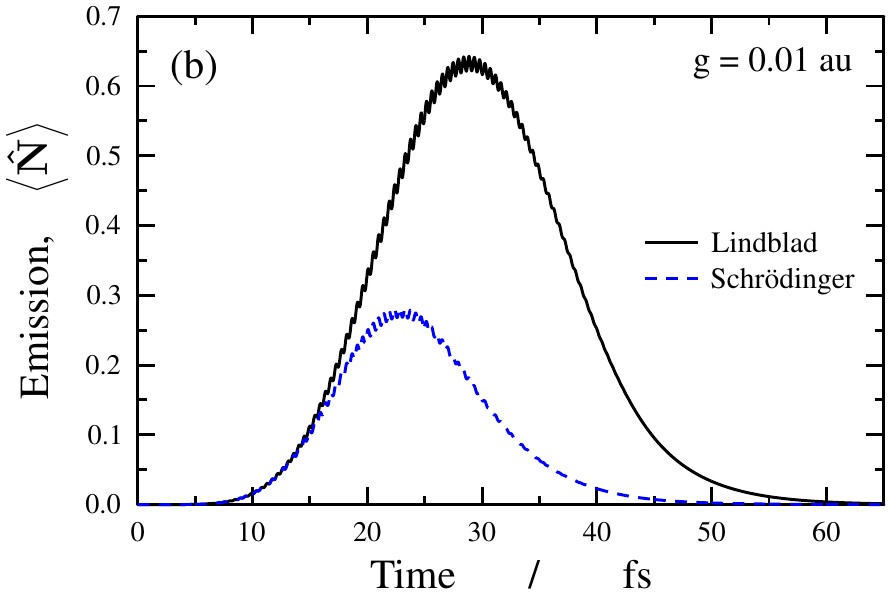}
\includegraphics[width=0.495\textwidth]{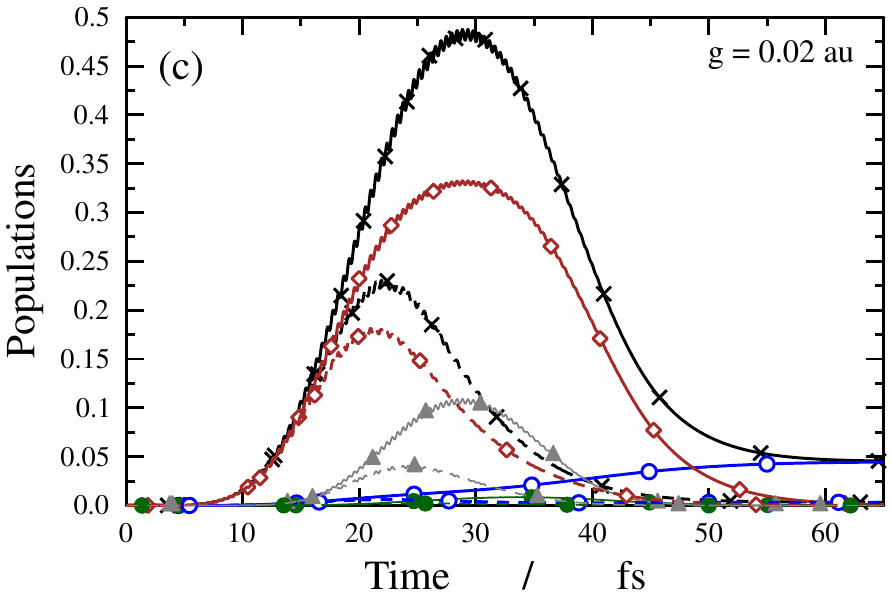}
\includegraphics[width=0.495\textwidth]{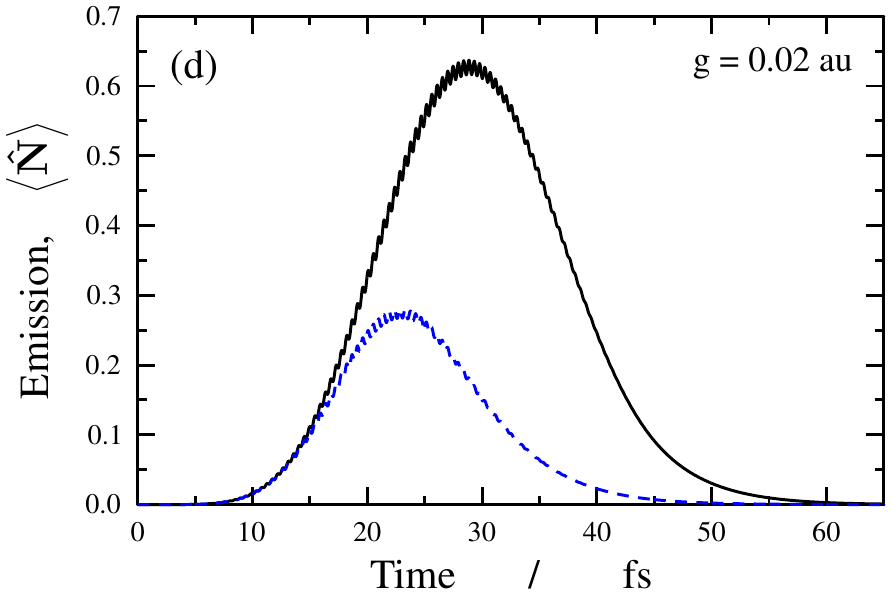}
\caption{\label{fig:UP_popem_3}
Populations of polaritonic states and emission curves
(the emission is proportional to the expectation value of the photon number operator $\hat{N}$)
for coupling strength values of $g=0.01~\textrm{au}$ (panels a and b)
and $g=0.02~\textrm{au}$ (panels c and d).
The cavity wavenumber and decay rate equal
$\omega_{\textrm{c}}=35744.8~\textrm{cm}^{-1}$ and   $\gamma_{\textrm{c}}=0.005~\textrm{au}$
(equivalent to a lifetime of $\tau = 4.8~\textrm{fs}$),
respectively.
Parameters of the pump laser pulse are chosen as
$\omega=36000~\textrm{cm}^{-1}$, $T=45~\textrm{fs}$ and 
$I=10^{12}~\textrm{W}/\textrm{cm}^{2}$.
The Lindblad results (solid lines) differ substantially from
their Schr\"odinger (TDSE) counterparts (dashed lines).}
\end{figure}

In this case, the laser and cavity wavenumbers are set to $\omega=36000~\textrm{cm}^{-1}$
and $\omega_{\textrm{c}}=35744.8~\textrm{cm}^{-1}$, respectively.
The corresponding population and emission results are presented in 
Figs. \ref{fig:UP_popem_1}, \ref{fig:UP_popem_2} and \ref{fig:UP_popem_3}.
In Fig. \ref{fig:UP_popem_1}, the remaining laser and cavity parameters are again chosen as
$T=15~\textrm{fs}$, $I=5\cdot10^{11}~\textrm{W}/\textrm{cm}^{2}$, $g=0.01~\textrm{au}$
as well as $\gamma_{\textrm{c}}=0.001~\textrm{au}$, $\gamma_{\textrm{c}}=0.0025~\textrm{au}$,
and $\gamma_{\textrm{c}}=0.005~\textrm{au}$. With these parameters, the pump pulse transfers population
dominantly to the photonic part of the 1UP state. We note that the current setup is related to 
case 2 of Ref. \onlinecite{22FaHaVi} where oscillations in the emission signal could be explained
by nonadiabatic population transfer between the 1UP and 1LP states.
The results shown in Fig. \ref{fig:UP_popem_1} are in line with those presented in
Fig. \ref{fig:LP_popem_1}. We can again observe that that the non-Hermitian TDSE method
is able to reproduce the emission signal and populations of excited polaritonic states
with high accuracy in this particular case.

Fig. \ref{fig:UP_popem_2} reveals trends similar to those observed in Fig. \ref{fig:LP_popem_2}.
Here we have employed the cavity and laser parameters $g=0.01~\textrm{au}$,
$\gamma_{\textrm{c}}=0.005~\textrm{au}$, $I=5\cdot10^{11}~\textrm{W}/\textrm{cm}^{2}$ or 
$I=10^{12}~\textrm{W}/\textrm{cm}^{2}$ and $T=30~\textrm{fs}$.
Besides the dominantly-populated 1UP state, the 2UP state also acquires some population
during laser excitation. The non-Hermitian TDSE description starts to break down in
Fig. \ref{fig:UP_popem_2}, manifesting in visible differences between the Lindblad and TDSE
populations and emissions. Indeed, the Lindblad approach again provides a much more efficient 
population transfer for longer pulses and higher intensities than the TDSE, which can be
explained by the arguments outlined in Section \ref{sec:results_case1}.

\begin{figure}
\includegraphics[width=0.495\textwidth]{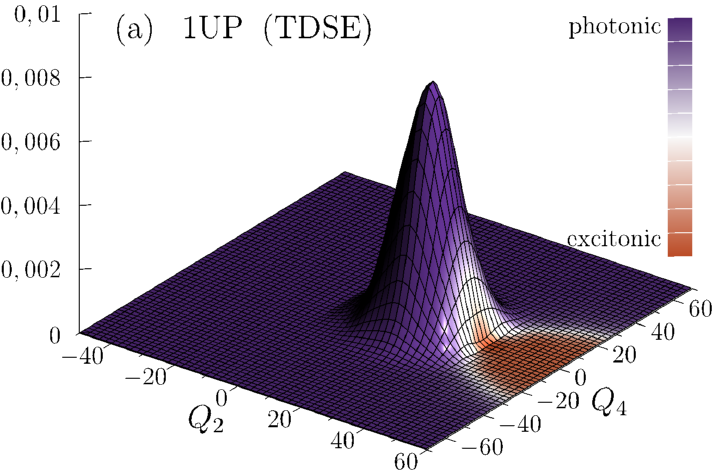}
\includegraphics[width=0.495\textwidth]{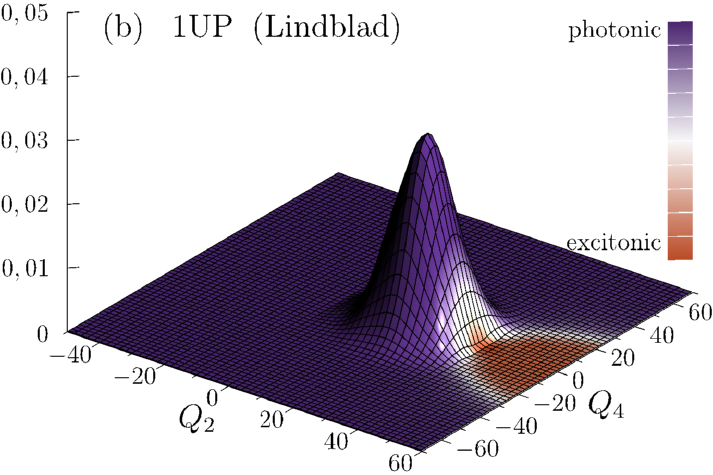}
\includegraphics[width=0.495\textwidth]{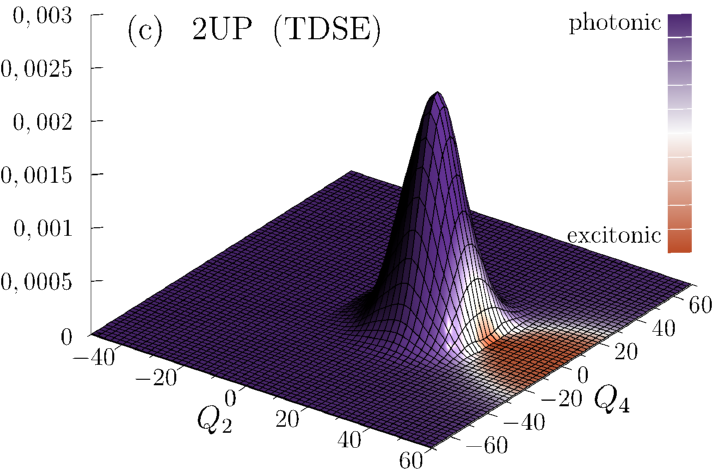}
\includegraphics[width=0.495\textwidth]{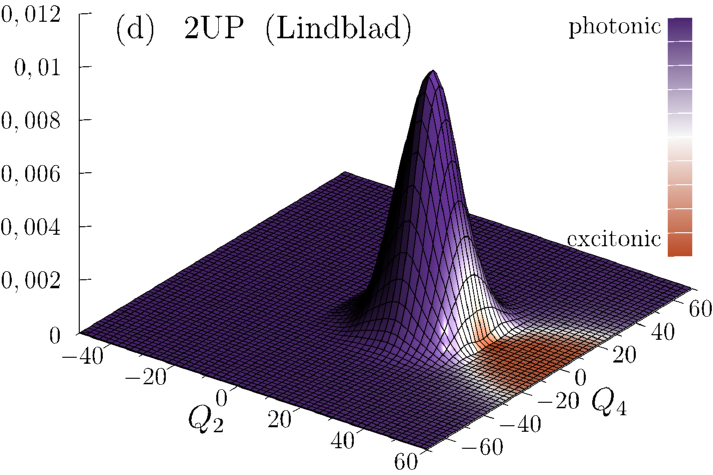}
\caption{\label{fig:UP_dens_1}
Probability density figures (obtained with the Lindblad and 
Schr\"odinger (TDSE) equations) at $t=30~\textrm{fs}$
for the 1UP (upper polaritonic, panels a and b) and 2UP 
(panels c and d) potential energy surfaces. 
The cavity wavenumber, coupling strength and decay rate equal
$\omega_{\textrm{c}}=35744.8~\textrm{cm}^{-1}$, $g=0.02~\textrm{au}$
and $\gamma_{\textrm{c}}=0.005~\textrm{au}$, respectively.
Parameters of the pump laser pulse are chosen as
$\omega=36000~\textrm{cm}^{-1}$, $T=45~\textrm{fs}$ and 
$I=10^{12}~\textrm{W}/\textrm{cm}^{2}$.
The laser pulse populates mainly the 1UP and 2UP states and the wave packet is localized in the photonic (purple) region in all
cases. Although the Lindblad and Schr\"odinger probability densities
have similar shapes, the norm of the Schr\"odinger probability densities are
definitely smaller than those of the Lindblad probability densities
(see the different scales along the $z$ axis).}
\end{figure}

\begin{figure}
\includegraphics[width=0.495\textwidth]{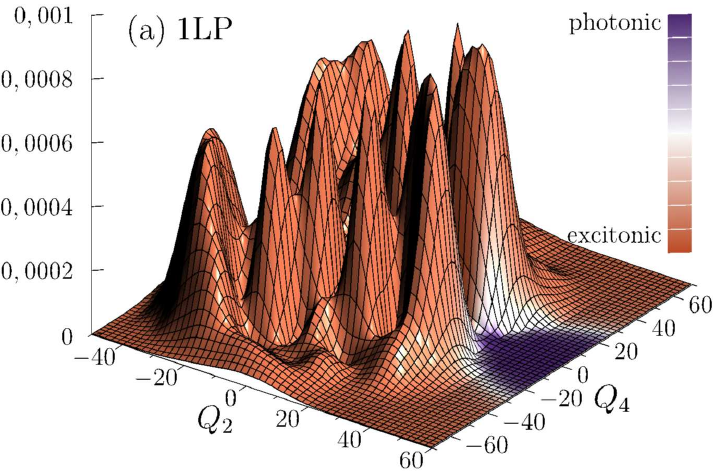}
\includegraphics[width=0.495\textwidth]{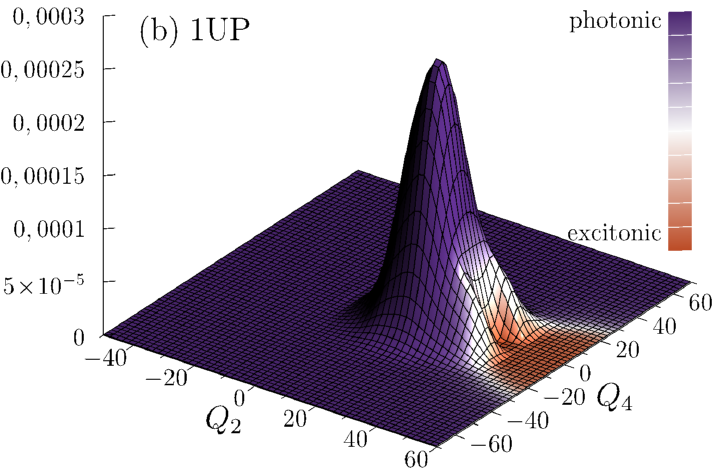}
\caption{\label{fig:UP_dens_2}
Probability density figures (obtained with the Lindblad equation)
at $t=60~\textrm{fs}$ for the 1LP (lower polaritonic, panel a)
and 1UP (upper polaritonic, panel b) potential energy surfaces. 
The cavity wavenumber, coupling strength and decay rate equal
$\omega_{\textrm{c}}=35744.8~\textrm{cm}^{-1}$, $g=0.02~\textrm{au}$
and $\gamma_{\textrm{c}}=0.005~\textrm{au}$, respectively.
Parameters of the pump laser pulse are chosen as
$\omega=36000~\textrm{cm}^{-1}$, $T=45~\textrm{fs}$ and 
$I=10^{12}~\textrm{W}/\textrm{cm}^{2}$. One can observe
nonadiabatic population transfer from the photonic (purple) region
of the initially-populated 1UP state to the excitonic (orange)
region of the 1LP state.}
\end{figure}

The differences between the Lindblad and TDSE methods are even more prominent in
Fig. \ref{fig:UP_popem_3}. Here, the pulse length has been increased to $T=45~\textrm{fs}$,
while other parameters has been set to $g=0.01~\textrm{au}$ or $g=0.02~\textrm{au}$,
$\gamma_{\textrm{c}}=0.005~\textrm{au}$ and $I=10^{12}~\textrm{W}/\textrm{cm}^{2}$.
For $g=0.02~\textrm{au}$ (see panel c of Fig. \ref{fig:UP_popem_3}) one can again notice
that some population is trapped in the 1LP state, but only in the Lindblad scheme. 
This is due to light-induced nonadiabatic behavior of the system. The 1LP and 1UP PESs form
a light-induced conical intersection (LICI) (see panel b of Fig. \ref{fig:PES}) which can serve
as a funnel for rapid nonadiabatic population transfer. 
This effect can be hardly seen for $g=0.01~\textrm{au}$, but becomes significant
for $g=0.02~\textrm{au}$. This finding is readily explained by the fact that light-induced nonadiabatic effects increase with the cavity coupling strength $g$.
Time-dependent probability densities shown in Figs. \ref{fig:UP_dens_1} and \ref{fig:UP_dens_2}
($g=0.02~\textrm{au}$) shed further light on the ultrafast dynamics through the LICI between
the 1UP and 1LP PESs.
Fig. \ref{fig:UP_dens_1} shows that although the Lindblad and TDSE probability densities have
similar shapes for the 1UP and 2UP PESs ($t=30~\textrm{fs}$), the TDSE probability densities have
considerably smaller norms than the Lindblad ones. This observation is verified by the different
scales of the $z$ axis in Fig. \ref{fig:UP_dens_1} and it is also in line with the population
curves given in Fig. \ref{fig:UP_popem_3}). Turning to Fig. \ref{fig:UP_dens_2} depicting 
Lindblad probability densities for $t=60~\textrm{fs}$, one can clearly see that dominant
part of the 1UP population is transferred to the excitonic part of the 1LP PES through the
LICI, which again quenches photon emission. This is possible only for the Lindblad master
equation as in the TDSE case the 1UP population does not persist until nonadiabatic population
transfer can set in.

Similarly to the Section \ref{sec:results_case1}, 
we have repeated the current TDSE calculations using the renormalized TDSE method.
The corresponding results are presented in Figs. \ref{fig:UP_popem_1_renorm}, 
\ref{fig:UP_popem_2_renorm} and \ref{fig:UP_popem_3_renorm} 
(see Appendix \ref{sec:appendixA}). The renormalized TDSE results show an almost perfect
agreement with their Lindblad counterparts, which is again due to the ground-state population
refill induced by the renormalization of the wave function.
The respective density matrix purities are depicted in Figs. 
\ref{fig:UP_purity_1}, \ref{fig:UP_purity_2} and \ref{fig:UP_purity_3} of Appendix
\ref{sec:appendixA}. In all cases considered in this section, the purity of the density
matrix equals one to a good approximation.
Results for the one-photon initial state used at the end of Section \ref{sec:results_case1}
are given in Appendix \ref{sec:appendixB} (see panels c and d of Figs. \ref{fig:special_1}
and \ref{fig:special_2} for $\omega_{\textrm{c}}=35744.8~\textrm{cm}^{-1}$).
Here we can again observe the breakdown of the renormalized TDSE method, which, together with 
the shape of the respective purity function (see the curve labeled as UP in Fig. 
\ref{fig:special_purity}), can be understood along the lines explained at the end of 
Section \ref{sec:results_case1}.

\section{Conclusions}
\label{sec:conclusions}
In this paper, the performance of the non-Hermitian time-dependent Schr\"odinger equation 
(TDSE) has been compared to that of the Lindblad-master-equation approach
for lossy plasmonic nanocavities.
As test system, a two-dimensional vibrational model of the four-atomic formaldehyde molecule
has been used. To carry out a thorough comparison of the two methods, several different 
cavity and laser pumping parameters have been applied. 

To summarize the findings of this paper we may say the following. 
(i) First of all, if the lower (1LP) and upper (1UP) polaritonic states in the
singly-excited subspace are driven by a laser pulse,
the TDSE with a non-Hermitian Hamiltonian 
(without renormalization of the wave function) is found to work well
in several situations (typically for lower laser intesities and shorter pulses) concerning
the excited-state populations and cavity emission signal.
However, Lindblad and TDSE populations of the ground polaritonic state always
differ since the TDSE method does not incorporate incoherent transitions which return
population from excited polaritonic states to the ground polaritonic state.
(ii) For larger laser intensities, or sufficiently long pulses (compared to the cavity 
photon lifetime), the TDSE results show visible or even substantial deviations from
their Lindblad counterparts, even for the populations of excited polaritonic states and
the cavity emission signal. A detailed analysis presented in Section \ref{sec:results} has
shown that the breakdown of the non-Hermitian TDSE method can be explained by the interplay
of ``upward'' laser excitation and ``downward'' incoherent transitions.
(iii) It is eye-catching that for all cases studied, the TDSE reproduces excellently all populations and emission signals for short times.
(iv) We have also demonstrated that the discrepancies between the Lindblad
and non-Hermitian TDSE results can be greatly reduced by renormalizing the wave function
at each time step during the numerical solution of the TDSE.

An important advantage of the non-Hermitian TDSE is that it enables the inclusion of
many more nuclear degrees of freedom (rotational and vibrational) in the quantum-dynamical
description due to its  lower computational cost compared to the Lindblad master equation.
Thus, the non-Hermitian TDSE, in the appropriate range of parameters
(and also in combination with wave function renormalization), 
can also offer a correct treatment of light-induced nonadiabatic properties of small
diatomic molecules in a lossy nanocavity.

\begin{acknowledgments}
The authors are indebted to NKFIH for funding (Grants No. K146096 and K138233).
Financial support by the Deutsche Forschungsgemeinschaft (DFG) (Grant No. CE 10/56-1)
is gratefully acknowledged. The work performed in Budapest received funding from 
the HUN-REN Hungarian Research Network.
This publication supports research performed within the COST Actions CA21101 
``Confined molecular systems: from a new generation of materials to the stars'' (COSY), and CA18222 ``Attosecond Chemistry'' (AttoChem),
funded by the European Cooperation in Science and Technology (COST),
and the PHYMOL (Physics, Accuracy and Machine 
Learning: Towards the Next Generation of Molecular Potentials) project,
funded mainly under the Horizon Europe scheme.
\end{acknowledgments}

\section*{Data availability} 
The data that support the findings of this study are available from the corresponding
author upon reasonable request.

\section*{Conflicts of interest}
The authors have no conflicts to disclose.

\appendix

\section{Renormalized Schr\"odinger (TDSE) results and density matrix purities}
\label{sec:appendixA}

\begin{figure}
\includegraphics[width=0.495\textwidth]{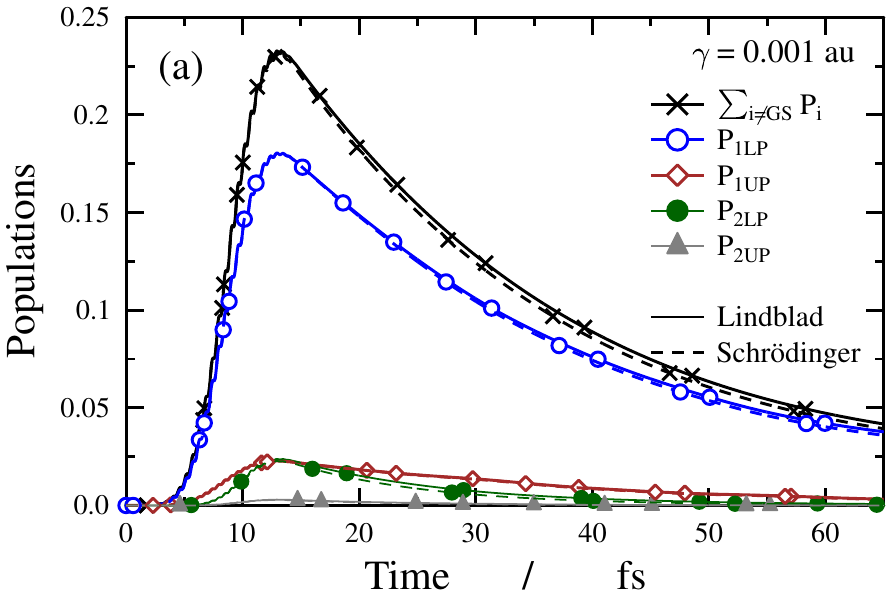}
\includegraphics[width=0.495\textwidth]{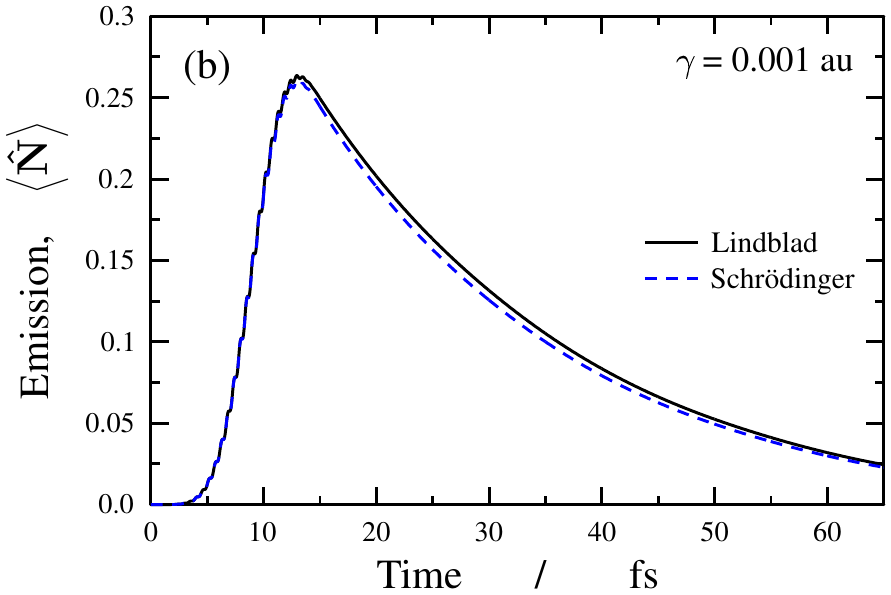}
\includegraphics[width=0.495\textwidth]{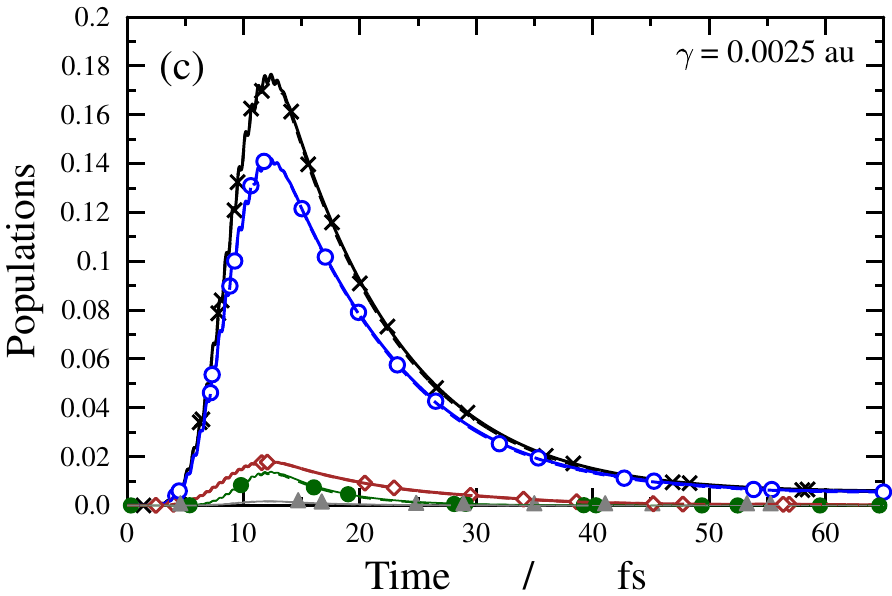}
\includegraphics[width=0.495\textwidth]{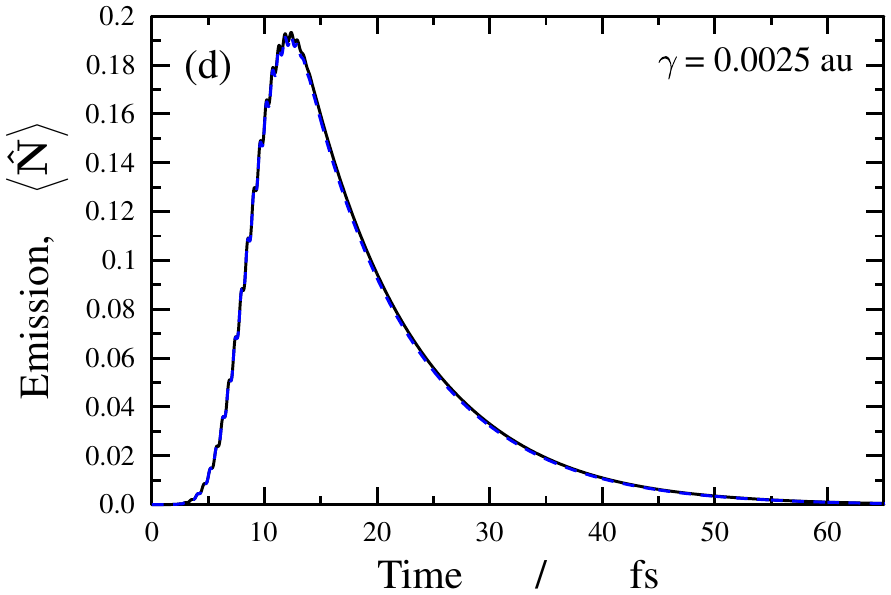}
\includegraphics[width=0.495\textwidth]{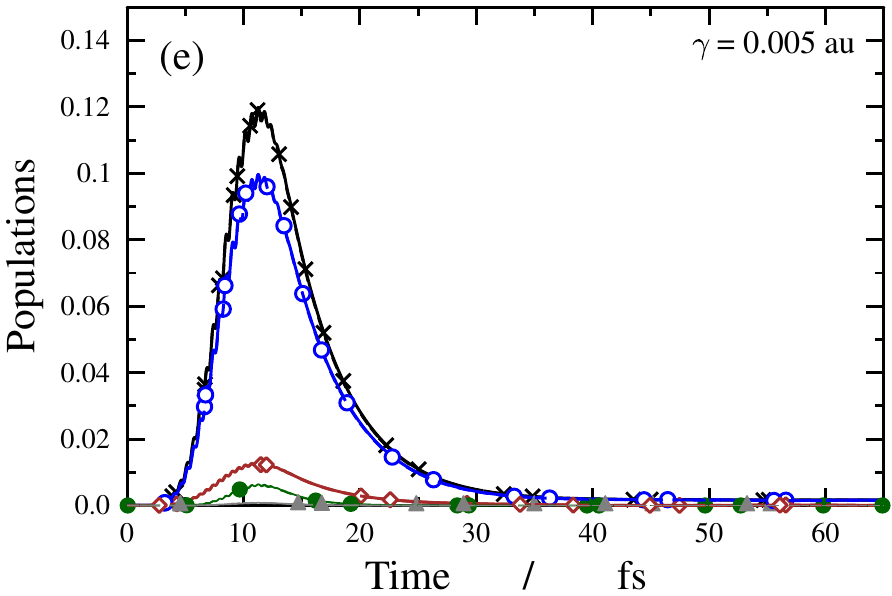}
\includegraphics[width=0.495\textwidth]{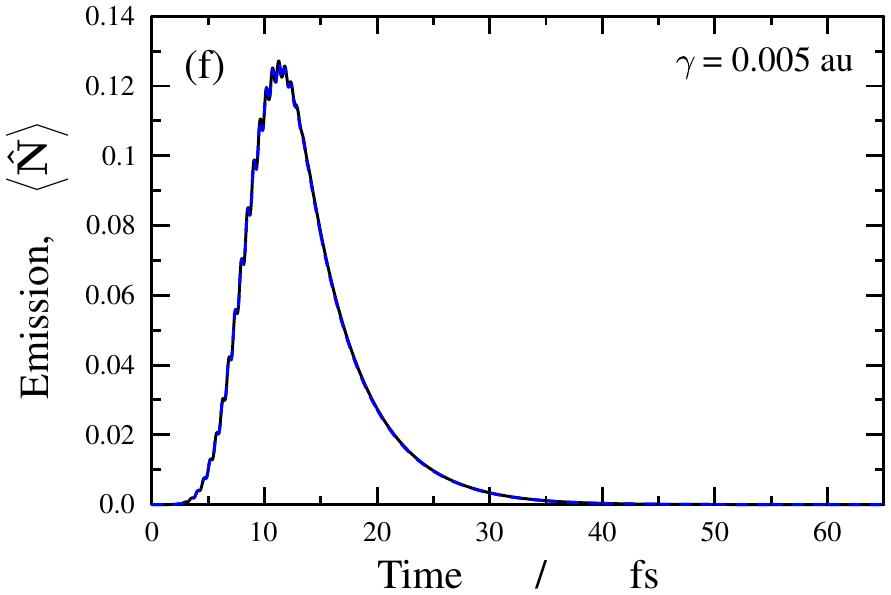}
\caption{\label{fig:LP_popem_1_renorm} 
Populations of polaritonic states and emission curves
(the emission is proportional to the expectation value of the photon number operator $\hat{N}$)
for cavity decay rates $\gamma_{\textrm{c}}=0.001~\textrm{au}$
(equivalent to a lifetime of $\tau = 24.2~\textrm{fs}$, panels a and b), 
$\gamma_{\textrm{c}}=0.0025~\textrm{au}$
($\tau = 9.7~\textrm{fs}$, panels c and d) and 
$\gamma_{\textrm{c}}=0.005~\textrm{au}$
($\tau = 4.8~\textrm{fs}$, panels e and f).
The Lindblad and renormalized Schr\"odinger (TDSE) results are depicted by solid 
and dashed lines, respectively.
The cavity wavenumber and coupling strength
equal $\omega_{\textrm{c}}=29957.2~\textrm{cm}^{-1}$
and $g=0.01~\textrm{au}$, respectively. Parameters of the pump
laser pulse are chosen as $\omega=30000~\textrm{cm}^{-1}$,
$T=15~\textrm{fs}$ and $I=5\cdot10^{11}~\textrm{W}/\textrm{cm}^{2}$.}
\end{figure}

\begin{figure}
\includegraphics[width=0.65\textwidth]{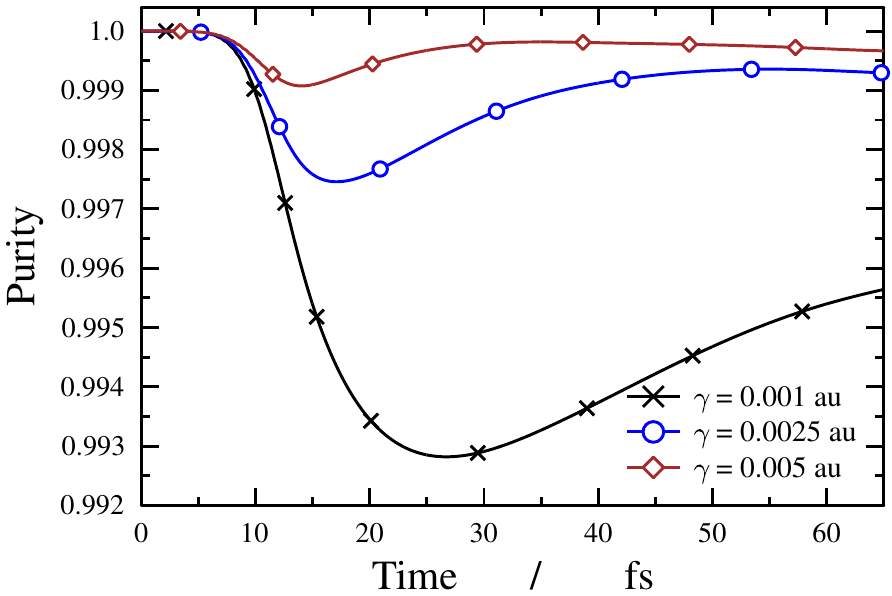}
\caption{\label{fig:LP_purity_1}
Purity of the density matrix ($\textrm{tr}(\hat{\rho}^2)$) as a function of time for 
cavity decay rates $\gamma_{\textrm{c}}=0.001~\textrm{au}$, 
$\gamma_{\textrm{c}}=0.0025~\textrm{au}$ and $\gamma_{\textrm{c}}=0.005~\textrm{au}$.
Unspecified cavity and laser parameters correspond to those applied in Figs. 
\ref{fig:LP_popem_1} and \ref{fig:LP_popem_1_renorm}.}
\end{figure}

\begin{figure}
\includegraphics[width=0.495\textwidth]{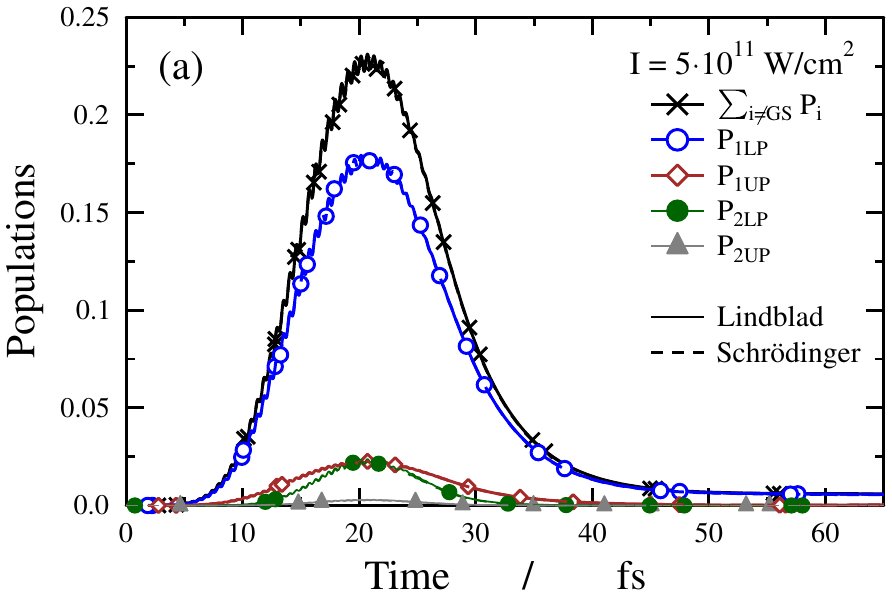}
\includegraphics[width=0.495\textwidth]{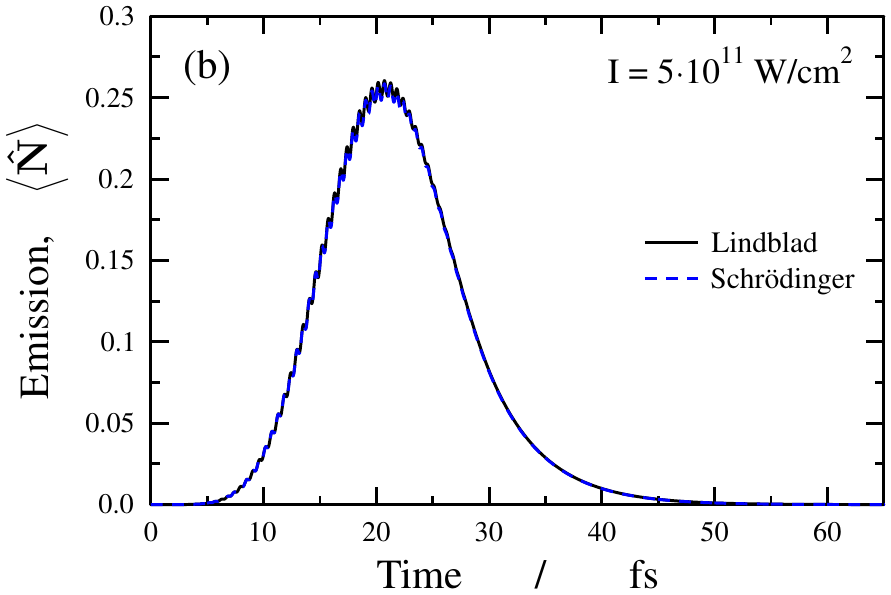}
\includegraphics[width=0.495\textwidth]{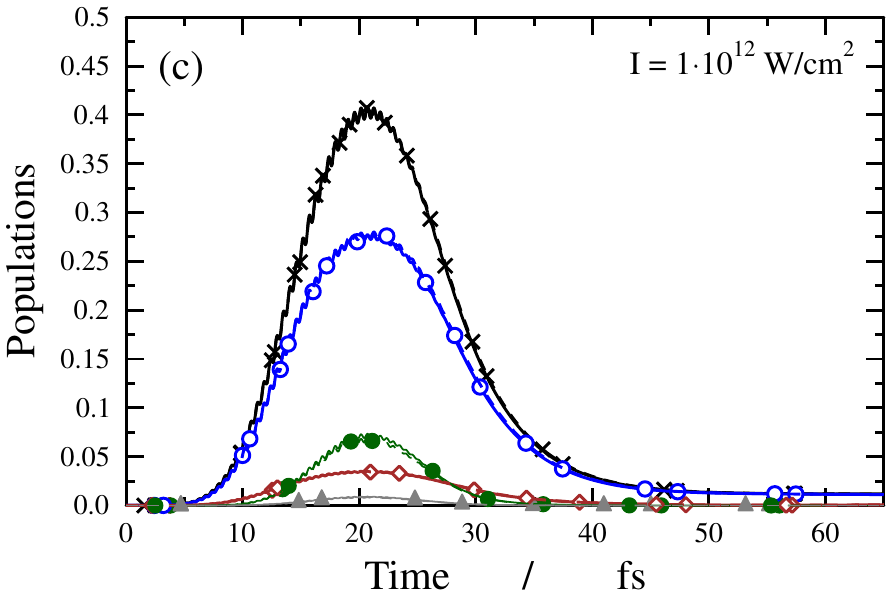}
\includegraphics[width=0.495\textwidth]{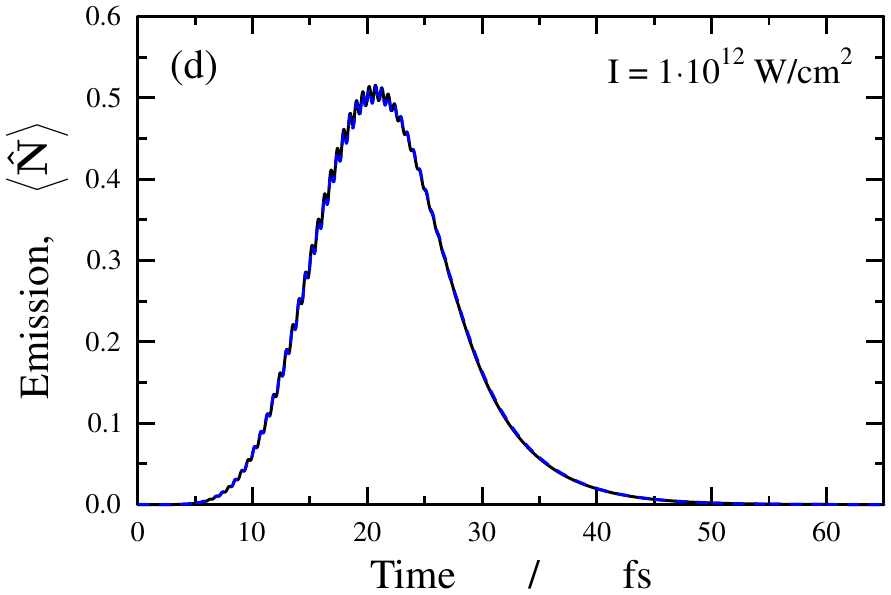}
\caption{\label{fig:LP_popem_2_renorm} 
Populations of polaritonic states and emission curves
(the emission is proportional to the expectation value of the photon number operator $\hat{N}$)
for laser intensities $I=5\cdot10^{11}~\textrm{W}/\textrm{cm}^{2}$
(panels a and b) and $I=10^{12}~\textrm{W}/\textrm{cm}^{2}$ (panels c and d). 
The Lindblad and renormalized Schr\"odinger (TDSE) results are depicted by solid 
and dashed lines, respectively.
The cavity wavenumber and coupling strength
equal $\omega_{\textrm{c}}=29957.2~\textrm{cm}^{-1}$
and $g=0.01~\textrm{au}$, respectively, while the cavity decay
rate is set to $\gamma_{\textrm{c}}=0.005~\textrm{au}$
(equivalent to a lifetime of $\tau = 4.8~\textrm{fs}$).
Other parameters of the pump laser pulse are chosen as
$\omega=30000~\textrm{cm}^{-1}$ and $T=30~\textrm{fs}$.}
\end{figure}

\begin{figure}
\includegraphics[width=0.65\textwidth]{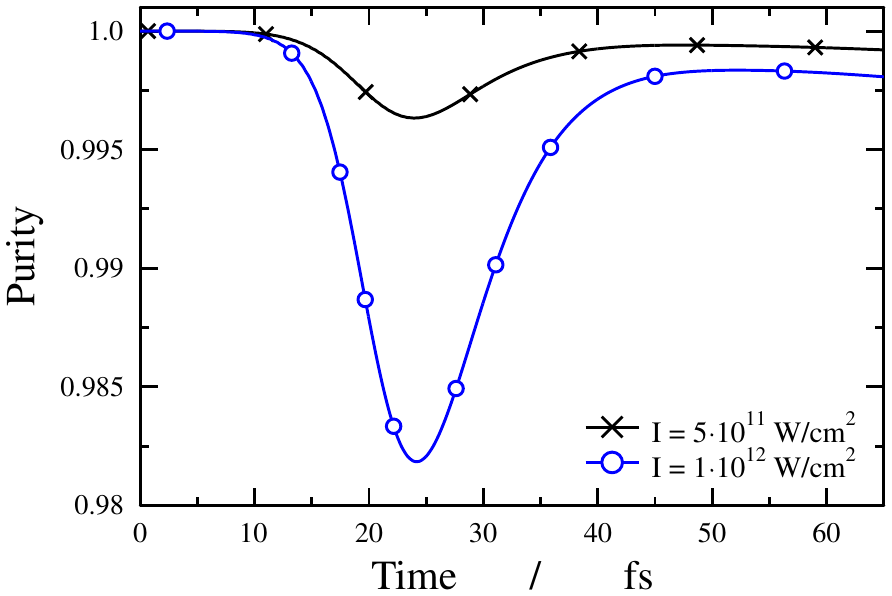}
\caption{\label{fig:LP_purity_2}
Purity of the density matrix ($\textrm{tr}(\hat{\rho}^2)$) as a function of time for 
laser intensities $I=5\cdot10^{11}~\textrm{W}/\textrm{cm}^{2}$ and
$I=10^{12}~\textrm{W}/\textrm{cm}^{2}$.
Unspecified cavity and laser parameters correspond to those applied in Figs. 
\ref{fig:LP_popem_2} and \ref{fig:LP_popem_2_renorm}.}
\end{figure}

\begin{figure}
\includegraphics[width=0.495\textwidth]{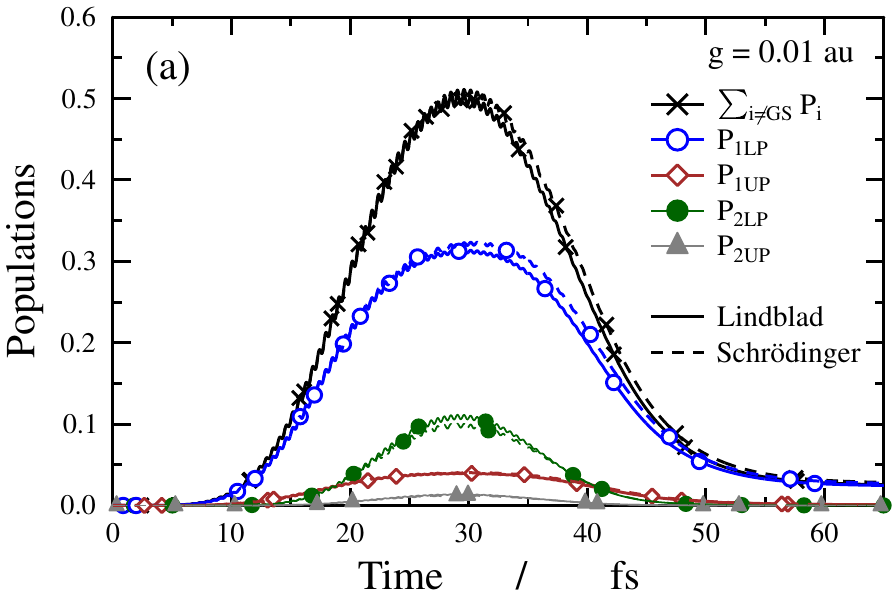}
\includegraphics[width=0.495\textwidth]{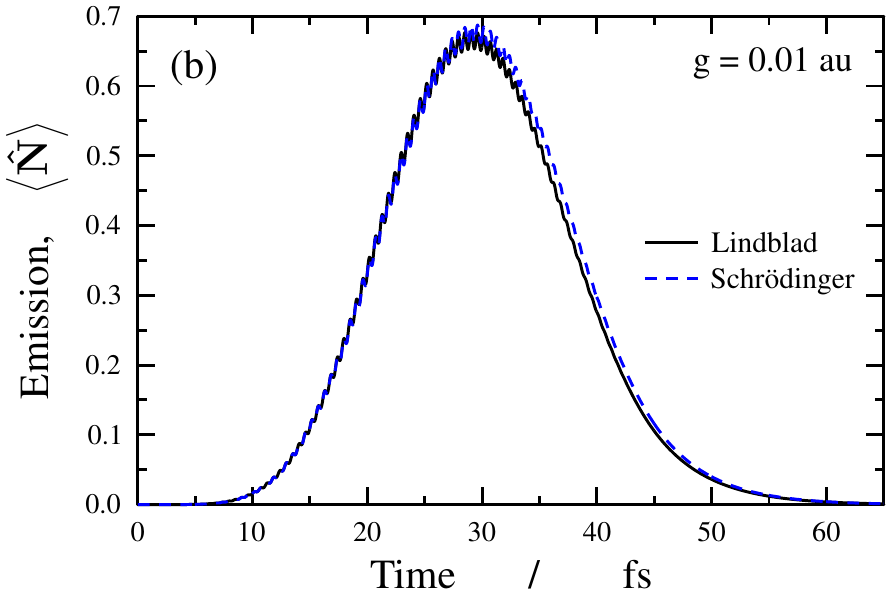}
\includegraphics[width=0.495\textwidth]{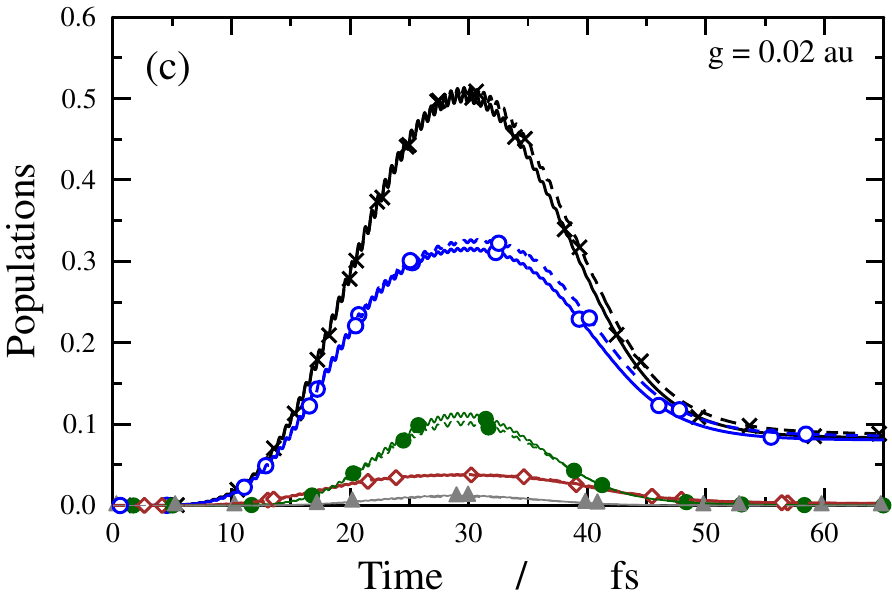}
\includegraphics[width=0.495\textwidth]{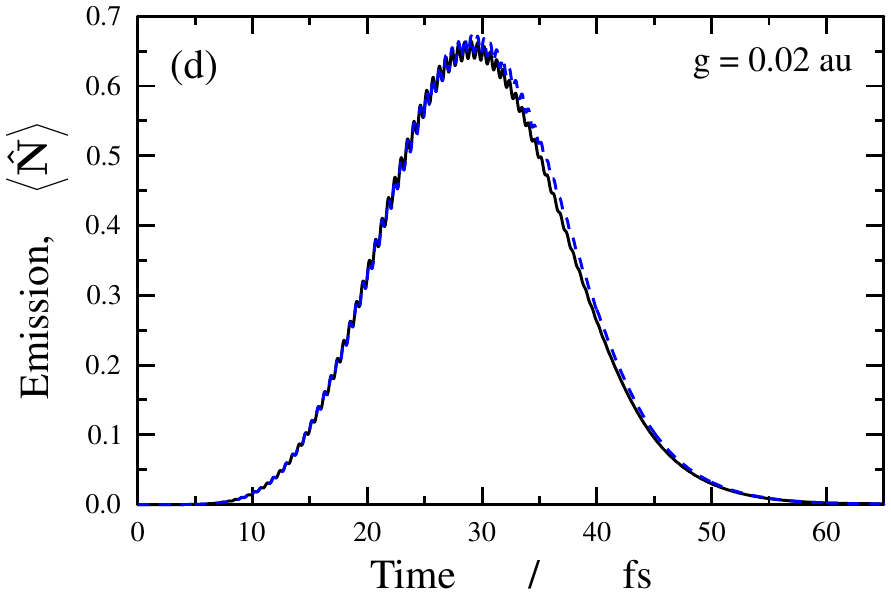}
\caption{\label{fig:LP_popem_3_renorm}
Populations of polaritonic states and emission curves (the emission is proportional to the
expectation value of the photon number operator $\hat{N}$) 
for coupling strength values of
$g=0.01~\textrm{au}$ (panels a and b) and $g=0.02~\textrm{au}$ (panels c and d).
The Lindblad and renormalized Schr\"odinger (TDSE) results are depicted by solid 
and dashed lines, respectively.
The cavity wavenumber and decay rate equal $\omega_{\textrm{c}}=29957.2~\textrm{cm}^{-1}$
and $\gamma_{\textrm{c}}=0.005~\textrm{au}$
(equivalent to a lifetime of $\tau = 4.8~\textrm{fs}$), respectively.
Parameters of the pump laser pulse are chosen as $\omega=30000~\textrm{cm}^{-1}$,
$T=45~\textrm{fs}$ and  $I=10^{12}~\textrm{W}/\textrm{cm}^{2}$.}
\end{figure}

\begin{figure}
\includegraphics[width=0.65\textwidth]{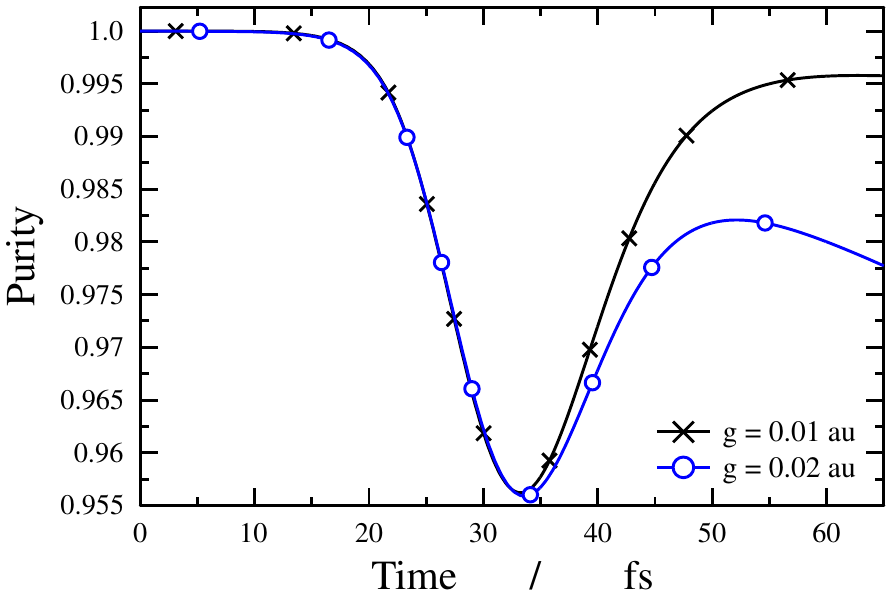}
\caption{\label{fig:LP_purity_3}
Purity of the density matrix ($\textrm{tr}(\hat{\rho}^2)$) as a function of time for 
coupling strength values of $g=0.01~\textrm{au}$ and $g=0.02~\textrm{au}$.
Unspecified cavity and laser parameters correspond to those applied in Figs. 
\ref{fig:LP_popem_3} and \ref{fig:LP_popem_3_renorm}.}
\end{figure}

\begin{figure}
\includegraphics[width=0.495\textwidth]{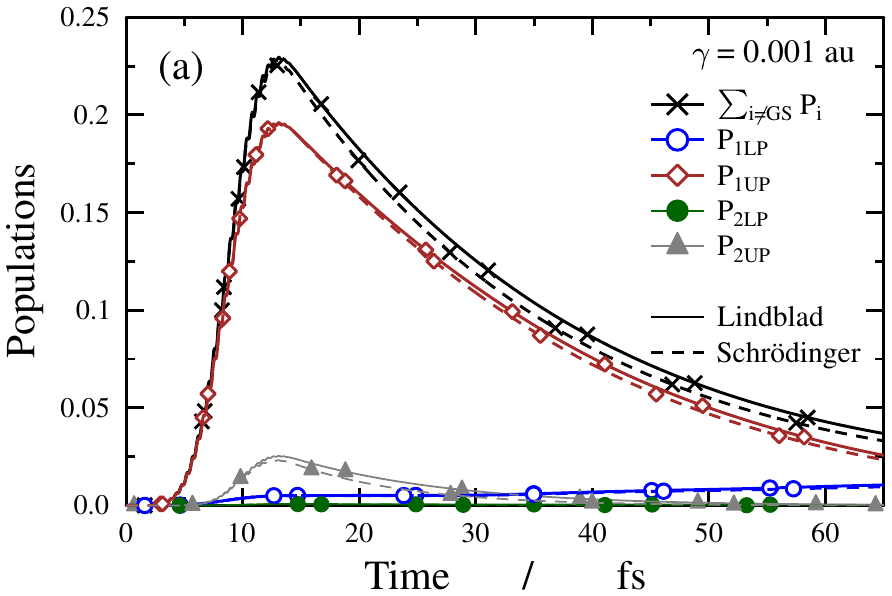}
\includegraphics[width=0.495\textwidth]{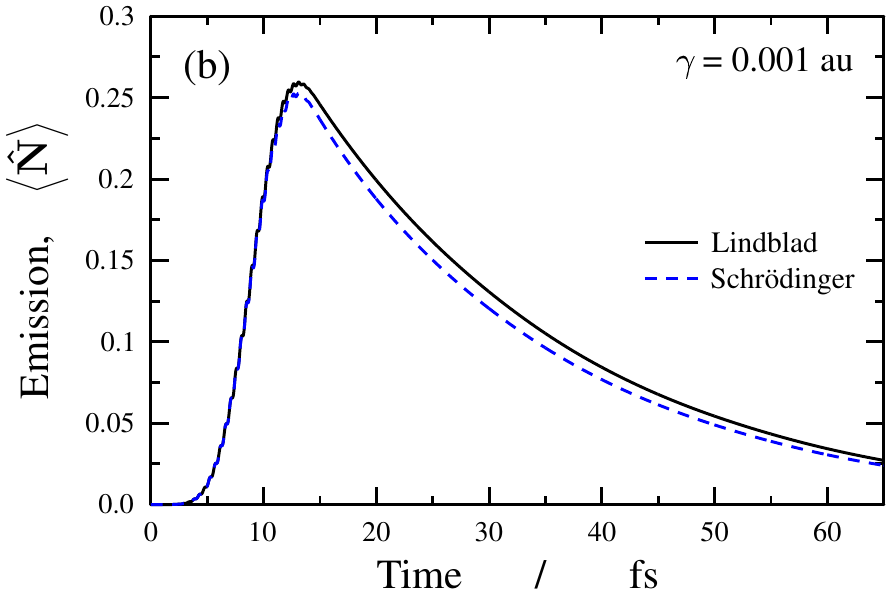}
\includegraphics[width=0.495\textwidth]{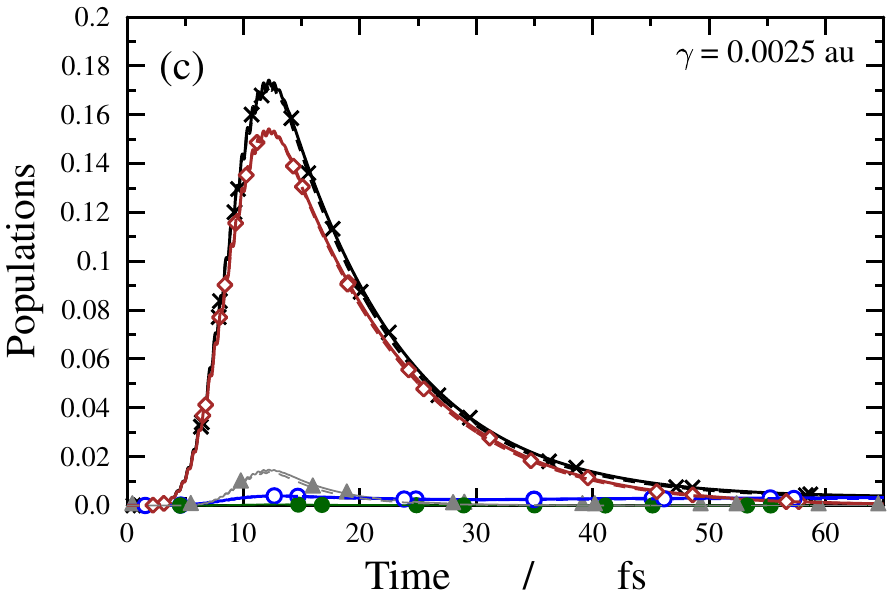}
\includegraphics[width=0.495\textwidth]{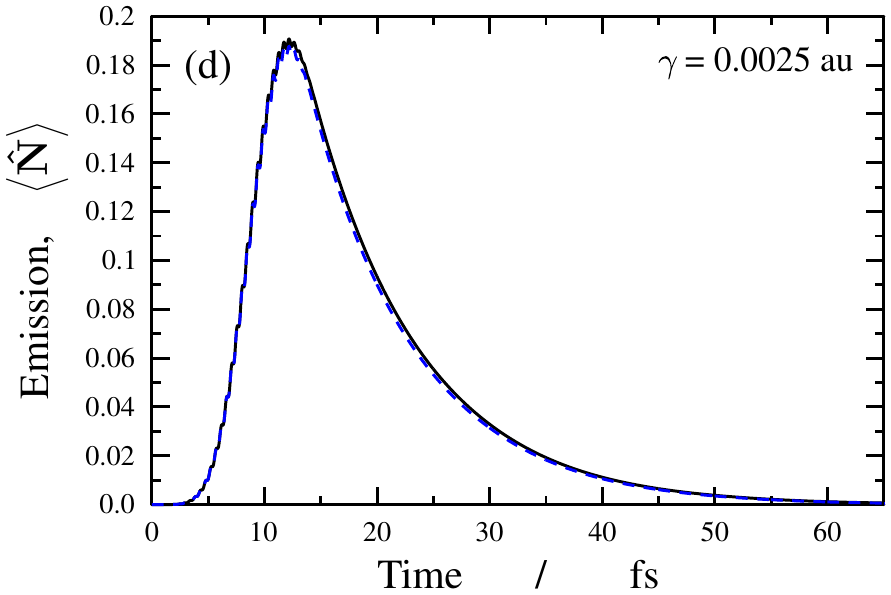}
\includegraphics[width=0.495\textwidth]{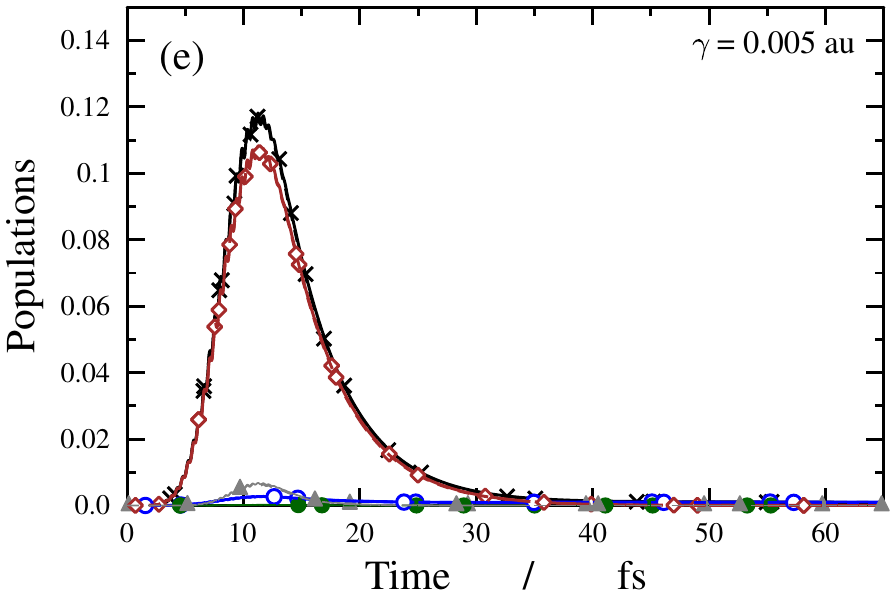}
\includegraphics[width=0.495\textwidth]{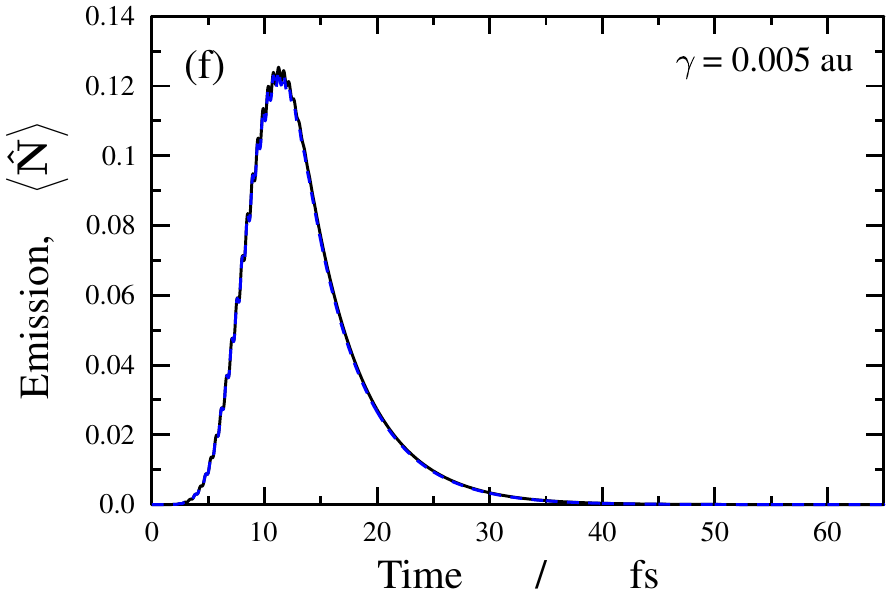}
\caption{\label{fig:UP_popem_1_renorm}
Populations of polaritonic states and emission curves
(the emission is proportional to the expectation value of the photon number operator $\hat{N}$)
for cavity decay rates $\gamma_{\textrm{c}}=0.001~\textrm{au}$
(equivalent to a lifetime of $\tau = 24.2~\textrm{fs}$, panels a and b), 
$\gamma_{\textrm{c}}=0.0025~\textrm{au}$
($\tau = 9.7~\textrm{fs}$, panels c and d) and 
$\gamma_{\textrm{c}}=0.005~\textrm{au}$
($\tau = 4.8~\textrm{fs}$, panels e and f).
The Lindblad and renormalized Schr\"odinger (TDSE) results are depicted by solid 
and dashed lines, respectively.
The cavity wavenumber and coupling strength
equal $\omega_{\textrm{c}}=35744.8~\textrm{cm}^{-1}$
and $g=0.01~\textrm{au}$, respectively. Parameters of the pump
laser pulse are chosen as $\omega=36000~\textrm{cm}^{-1}$,
$T=15~\textrm{fs}$ and $I=5\cdot10^{11}~\textrm{W}/\textrm{cm}^{2}$.}
\end{figure}

\begin{figure}
\includegraphics[width=0.65\textwidth]{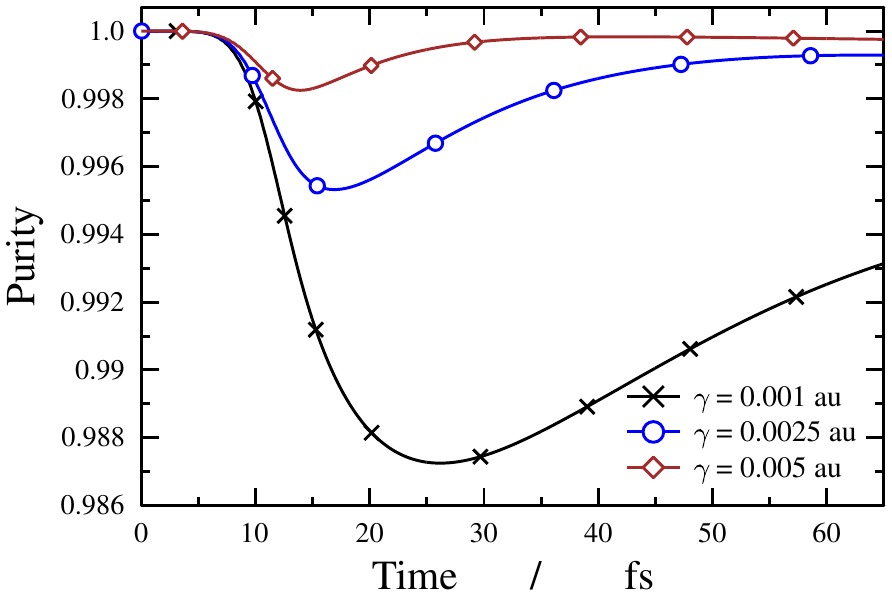}
\caption{\label{fig:UP_purity_1}
Purity of the density matrix ($\textrm{tr}(\hat{\rho}^2)$) as a function of time for 
cavity decay rates $\gamma_{\textrm{c}}=0.001~\textrm{au}$, 
$\gamma_{\textrm{c}}=0.0025~\textrm{au}$ and $\gamma_{\textrm{c}}=0.005~\textrm{au}$.
Unspecified cavity and laser parameters correspond to those applied in Figs. 
\ref{fig:UP_popem_1} and \ref{fig:UP_popem_1_renorm}.}
\end{figure}

\begin{figure}
\includegraphics[width=0.495\textwidth]{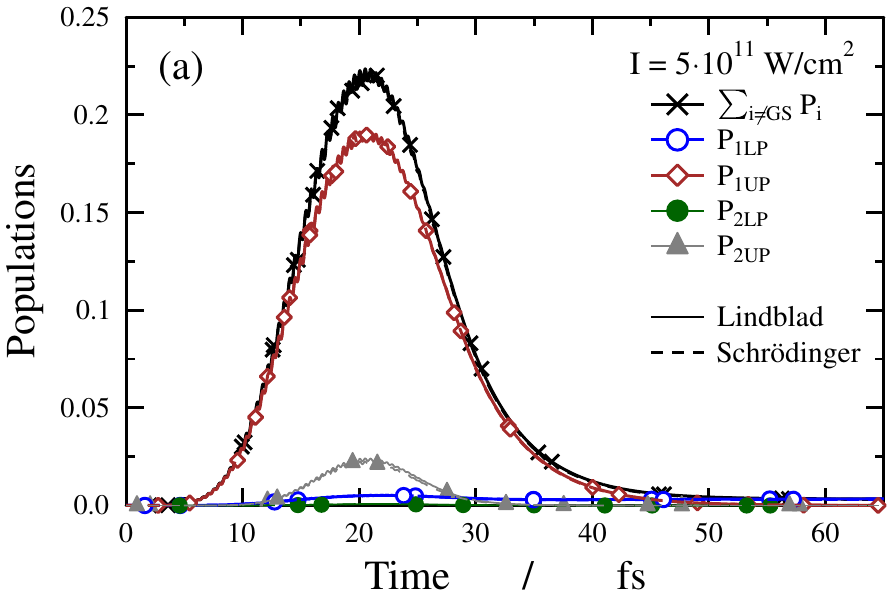}
\includegraphics[width=0.495\textwidth]{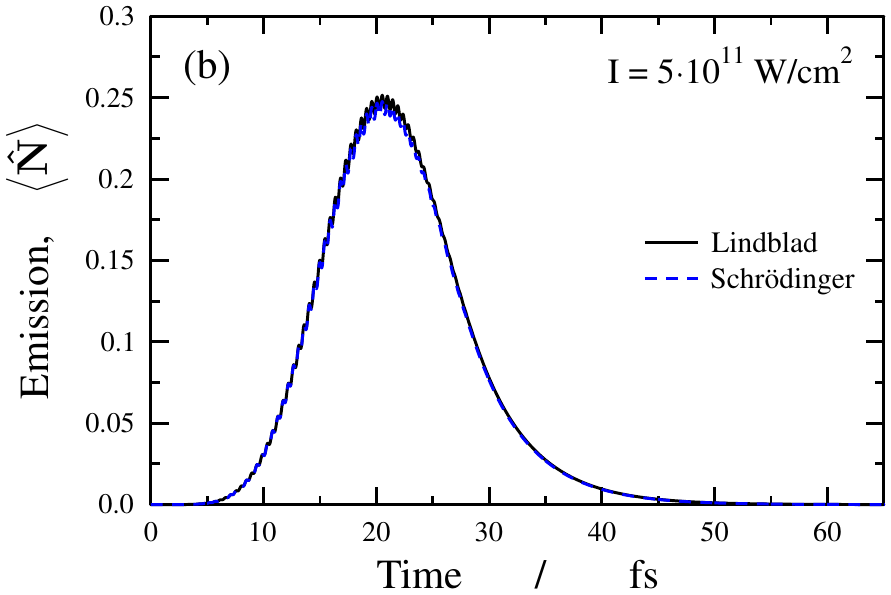}
\includegraphics[width=0.495\textwidth]{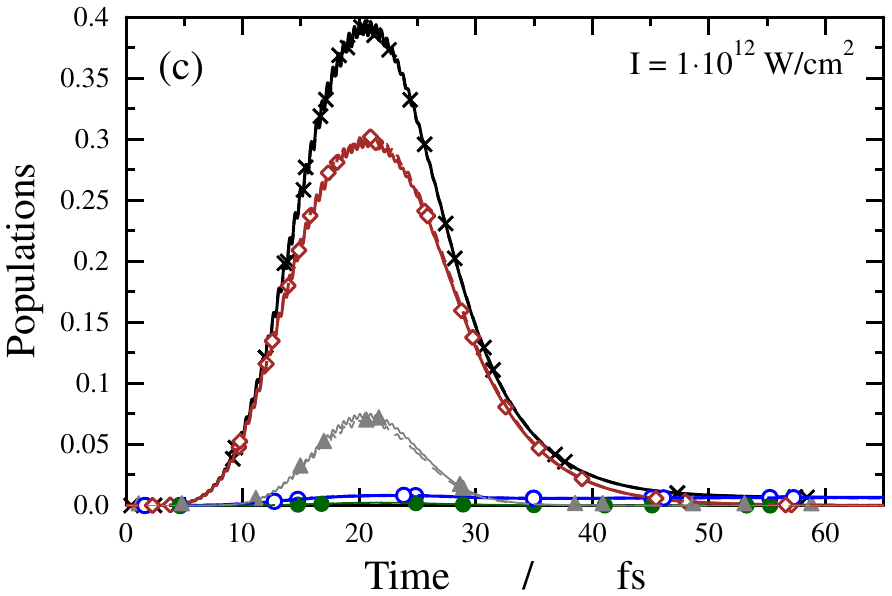}
\includegraphics[width=0.495\textwidth]{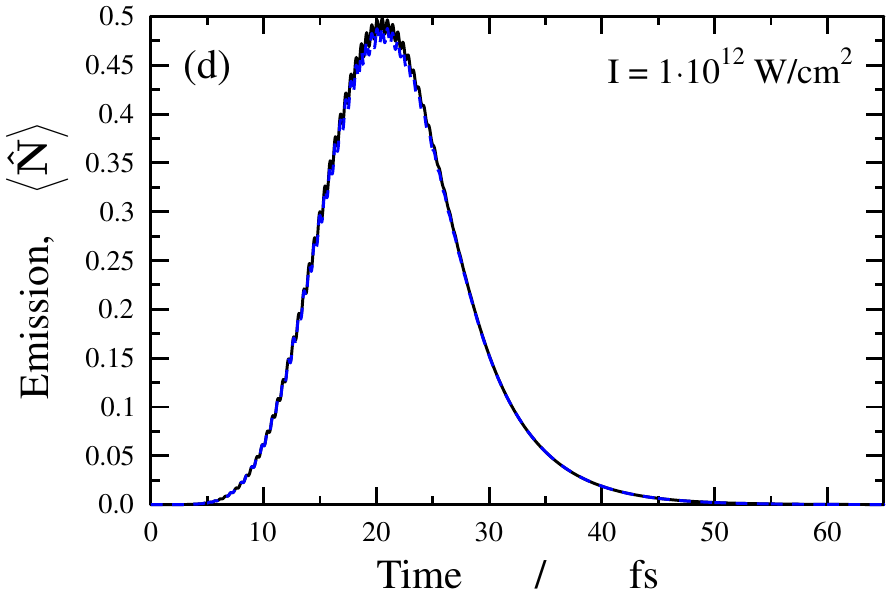}
\caption{\label{fig:UP_popem_2_renorm}
Populations of polaritonic states and emission curves 
(the emission is proportional to the expectation value of the photon number operator $\hat{N}$)
for laser intensities $I=5\cdot10^{11}~\textrm{W}/\textrm{cm}^{2}$
(panels a and b) and $I=10^{12}~\textrm{W}/\textrm{cm}^{2}$ (panels c and d).
The Lindblad and renormalized Schr\"odinger (TDSE) results are depicted by solid 
and dashed lines, respectively.
The cavity wavenumber and coupling strength
equal $\omega_{\textrm{c}}=35744.8~\textrm{cm}^{-1}$
and $g=0.01~\textrm{au}$, respectively, while the cavity decay
rate is set to $\gamma_{\textrm{c}}=0.005~\textrm{au}$
(equivalent to a lifetime of $\tau = 4.8~\textrm{fs}$).
Other parameters of the pump laser pulse are chosen as
$\omega=36000~\textrm{cm}^{-1}$ and $T=30~\textrm{fs}$.}
\end{figure}

\begin{figure}
\includegraphics[width=0.65\textwidth]{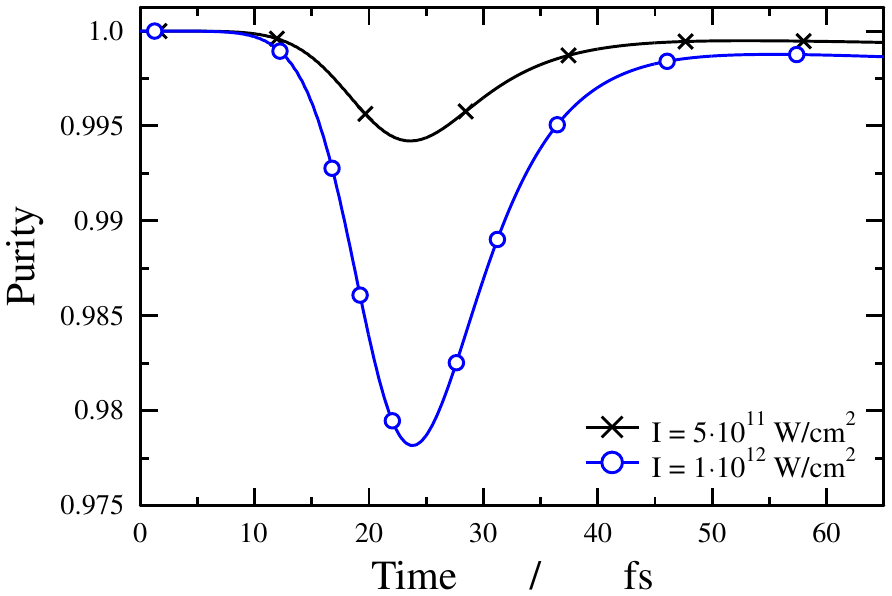}
\caption{\label{fig:UP_purity_2}
Purity of the density matrix ($\textrm{tr}(\hat{\rho}^2)$) as a function of time for 
laser intensities $I=5\cdot10^{11}~\textrm{W}/\textrm{cm}^{2}$ and
$I=10^{12}~\textrm{W}/\textrm{cm}^{2}$.
Unspecified cavity and laser parameters correspond to those applied in Figs. 
\ref{fig:UP_popem_2} and \ref{fig:UP_popem_2_renorm}.}
\end{figure}

\begin{figure}
\includegraphics[width=0.495\textwidth]{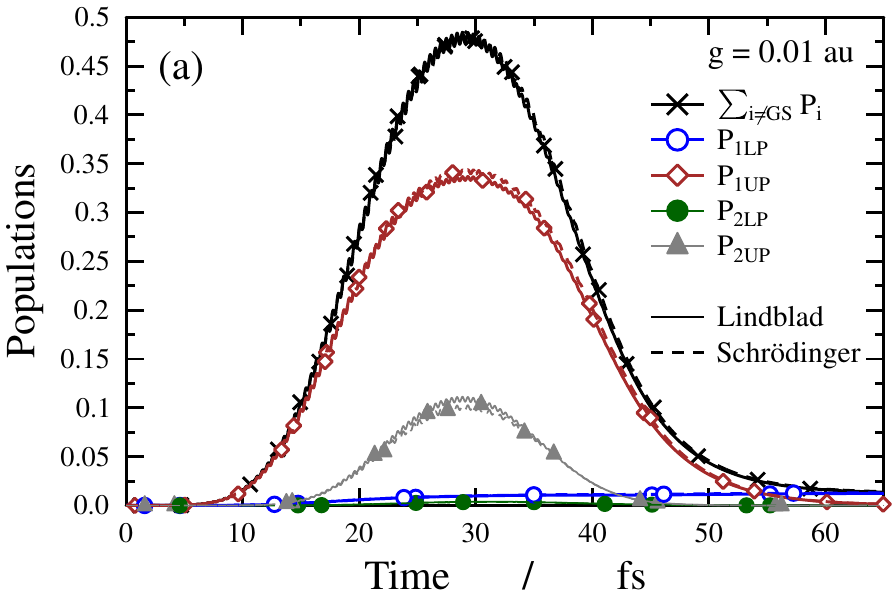}
\includegraphics[width=0.495\textwidth]{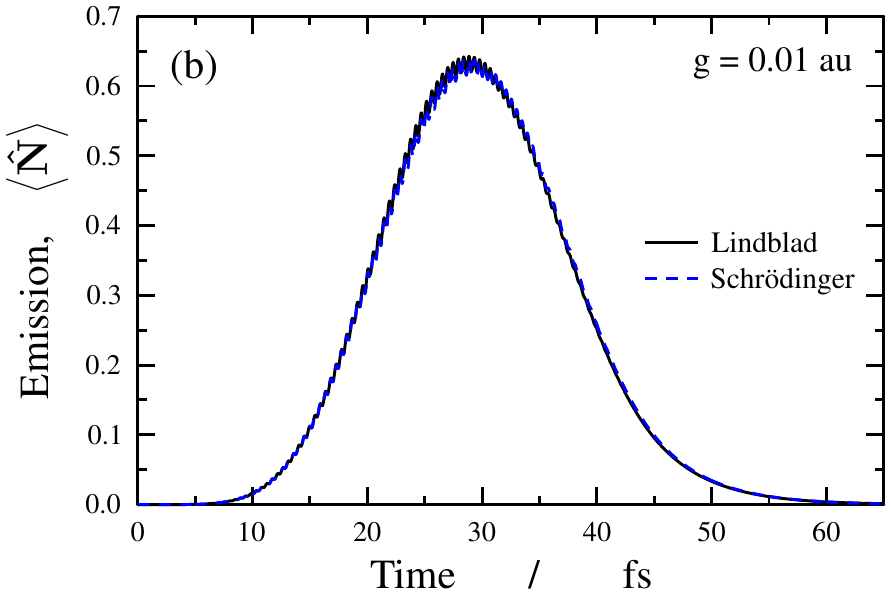}
\includegraphics[width=0.495\textwidth]{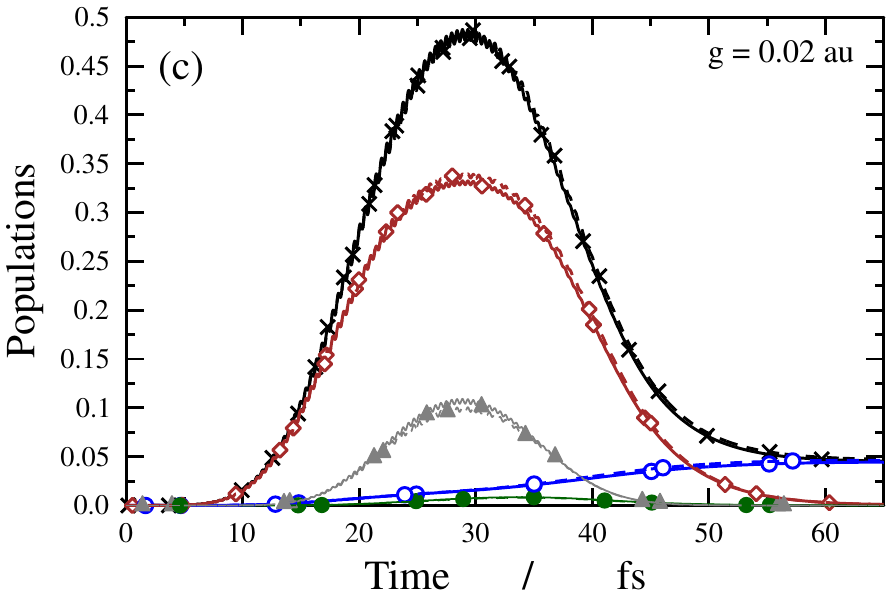}
\includegraphics[width=0.495\textwidth]{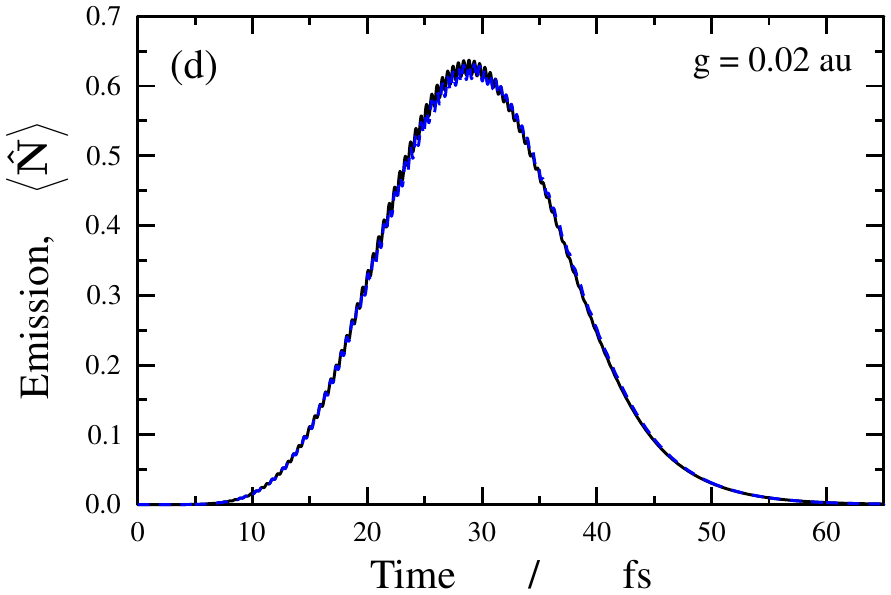}
\caption{\label{fig:UP_popem_3_renorm}
Populations of polaritonic states and emission curves
(the emission is proportional to the expectation value of the photon number operator $\hat{N}$)
for coupling strength values of $g=0.01~\textrm{au}$ (panels a and b)
and $g=0.02~\textrm{au}$ (panels c and d).
The Lindblad and renormalized Schr\"odinger (TDSE) results are depicted by solid 
and dashed lines, respectively.
The cavity wavenumber and decay rate equal
$\omega_{\textrm{c}}=35744.8~\textrm{cm}^{-1}$ and   $\gamma_{\textrm{c}}=0.005~\textrm{au}$
(equivalent to a lifetime of $\tau = 4.8~\textrm{fs}$), respectively.
Parameters of the pump laser pulse are chosen as
$\omega=36000~\textrm{cm}^{-1}$, $T=45~\textrm{fs}$ and 
$I=10^{12}~\textrm{W}/\textrm{cm}^{2}$.}
\end{figure}

\begin{figure}
\includegraphics[width=0.65\textwidth]{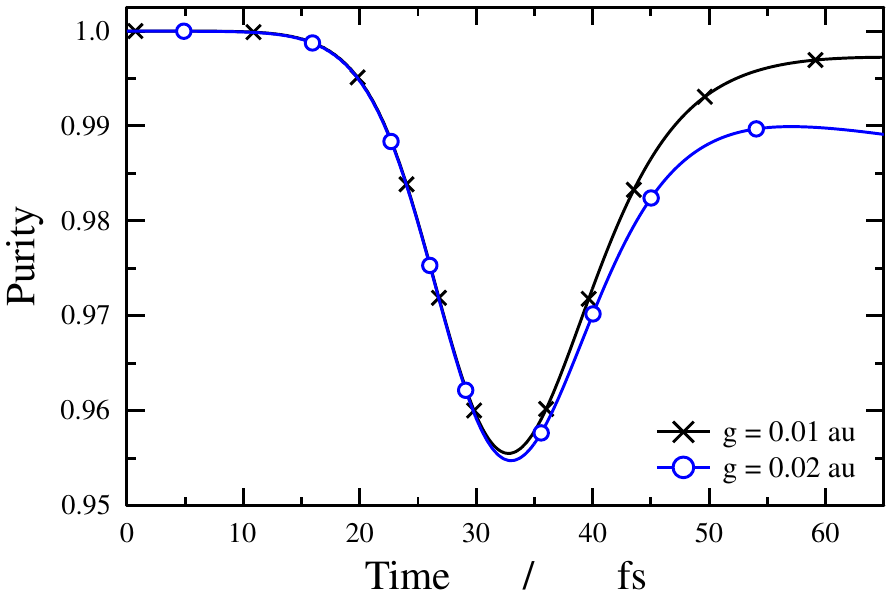}
\caption{\label{fig:UP_purity_3}
Purity of the density matrix ($\textrm{tr}(\hat{\rho}^2)$) as a function of time for 
coupling strength values of $g=0.01~\textrm{au}$ and $g=0.02~\textrm{au}$.
Unspecified cavity and laser parameters correspond to those applied in Figs. 
\ref{fig:UP_popem_3} and \ref{fig:UP_popem_3_renorm}.}
\end{figure}

Figs. \ref{fig:LP_popem_1_renorm}, 
\ref{fig:LP_popem_2_renorm}, \ref{fig:LP_popem_3_renorm}
and Figs. \ref{fig:UP_popem_1_renorm},
\ref{fig:UP_popem_2_renorm} and \ref{fig:UP_popem_3_renorm}
show population and emission results obtained with the 
Lindblad and renormalized TDSE methods using the parameters
of Section \ref{sec:results_case1} and
Section \ref{sec:results_case2}, respectively.
The corresponding density matrix purities 
($\textrm{tr}(\hat{\rho}^2)$)
are depicted in Figs. \ref{fig:LP_purity_1}, 
\ref{fig:LP_purity_2}, \ref{fig:LP_purity_3}
and Figs. \ref{fig:UP_purity_1}, 
\ref{fig:UP_purity_2}, \ref{fig:UP_purity_3}.
In all cases, the Lindblad results show an excellent
agreement with their counterparts obtained with the
renormalized TDSE method. Moreover, the purities equal
one at $t=0$ and remain close to one for all times
shown in the figures.

\clearpage
\section{Reference results with a special initial state}
\label{sec:appendixB}

\begin{figure}
\includegraphics[width=0.495\textwidth]{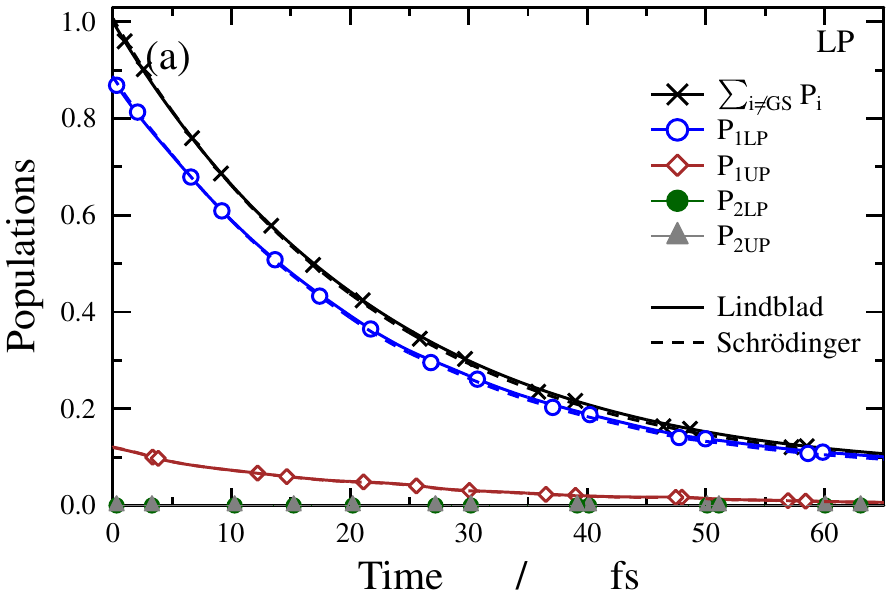}
\includegraphics[width=0.495\textwidth]{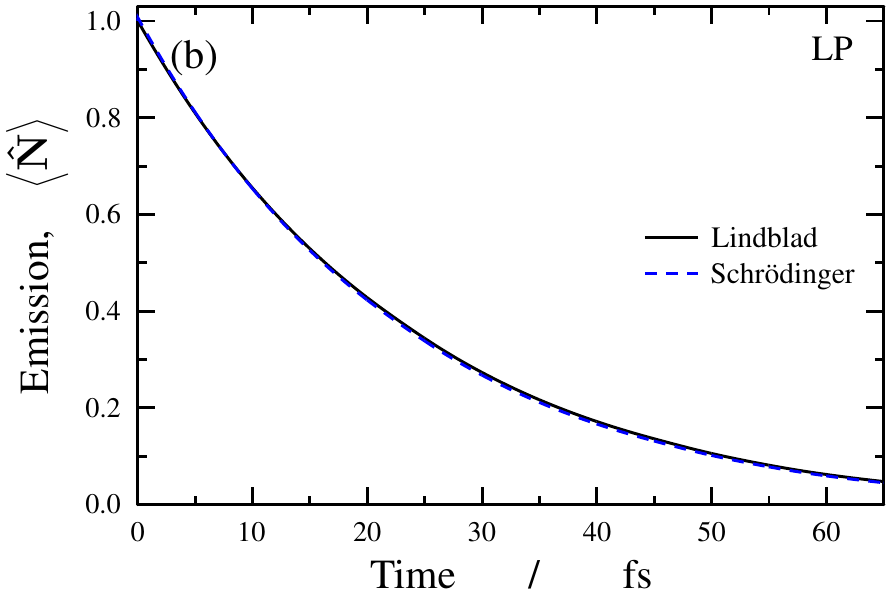}
\includegraphics[width=0.495\textwidth]{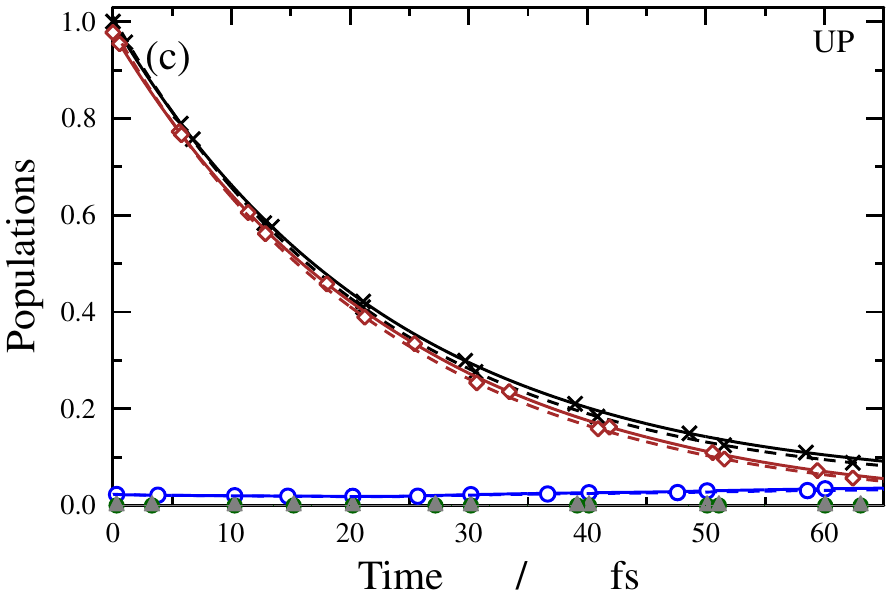}
\includegraphics[width=0.495\textwidth]{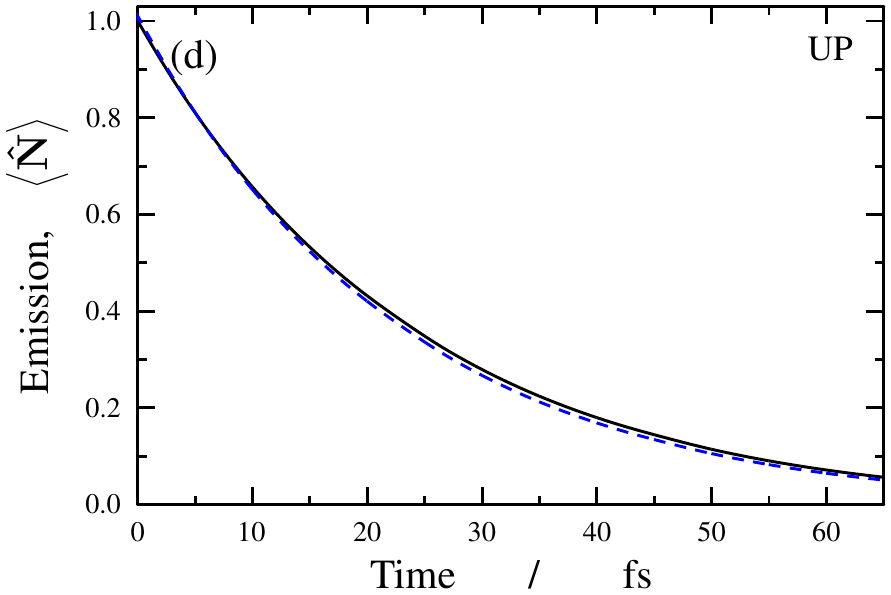}
\caption{\label{fig:special_1}
Populations of polaritonic states and emission curves (the emission is proportional 
to the expectation value of the photon number operator $\hat{N}$).
The initial state is chosen as follows: molecule in the vibrational ground state of the 
ground (X) electronic state with 1 photon in the cavity mode.
The Lindblad and Schr\"odinger (TDSE) results are depicted by solid 
and dashed lines, respectively. The cavity wavenumber is set to
$\omega_{\textrm{c}}=29957.2~\textrm{cm}^{-1}$ (panels a and b) or
$\omega_{\textrm{c}}=35744.8~\textrm{cm}^{-1}$ (panels c and d).
The coupling strength and cavity decay rate equal $g=0.01~\textrm{au}$ and
$\gamma_{\textrm{c}}=0.001~\textrm{au}$, respectively.}
\end{figure}

\begin{figure}
\includegraphics[width=0.495\textwidth]{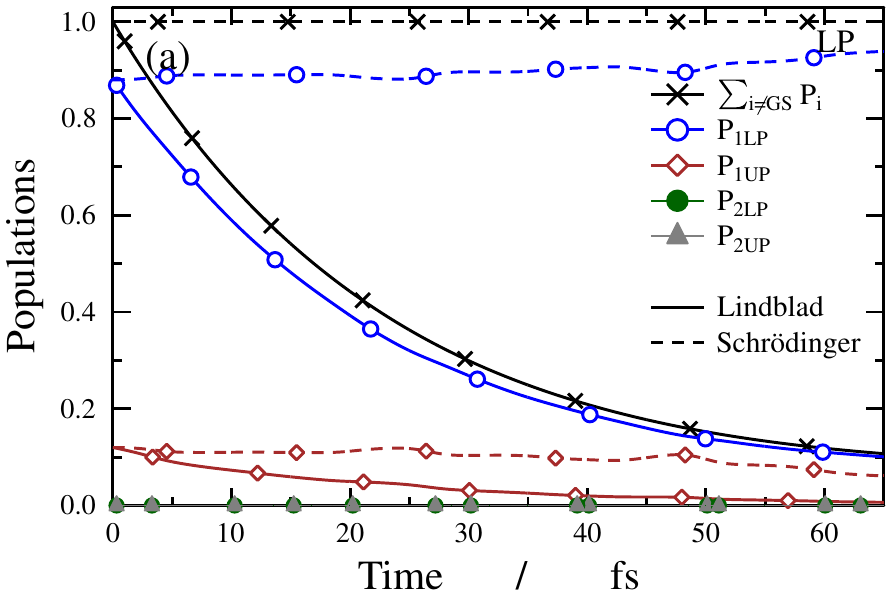}
\includegraphics[width=0.495\textwidth]{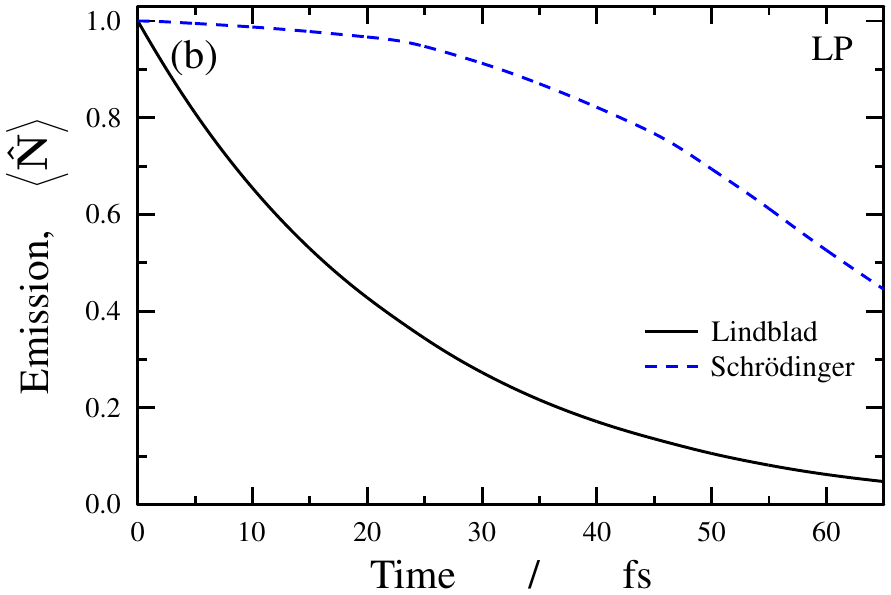}
\includegraphics[width=0.495\textwidth]{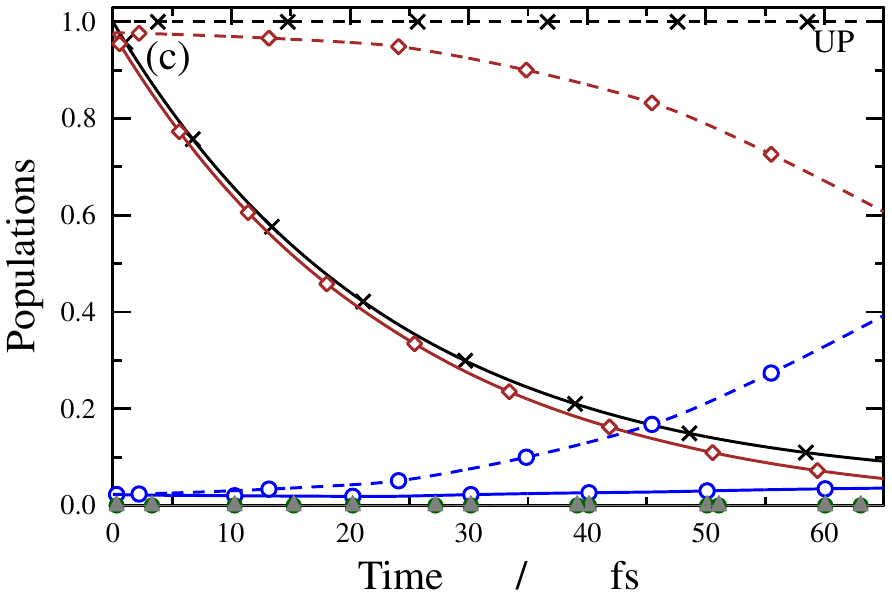}
\includegraphics[width=0.495\textwidth]{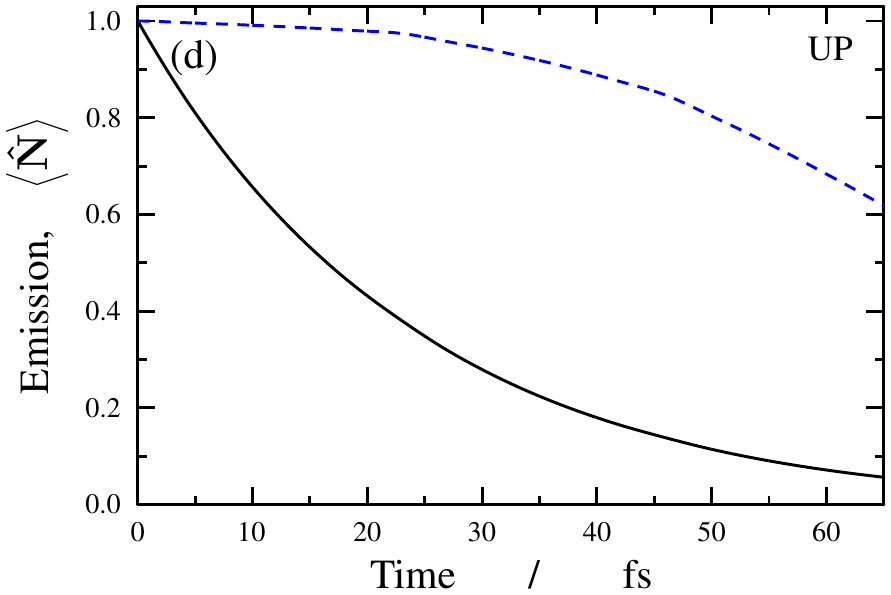}
\caption{\label{fig:special_2}
Populations of polaritonic states and emission curves (the emission is proportional 
to the expectation value of the photon number operator $\hat{N}$).
The initial state is chosen as follows: molecule in the vibrational ground state of the 
ground (X) electronic state with 1 photon in the cavity mode.
The Lindblad and renormalized Schr\"odinger (TDSE) results are depicted by solid 
and dashed lines, respectively. The cavity wavenumber is set to
$\omega_{\textrm{c}}=29957.2~\textrm{cm}^{-1}$ (panels a and b) or
$\omega_{\textrm{c}}=35744.8~\textrm{cm}^{-1}$ (panels c and d).
The coupling strength and cavity decay rate equal $g=0.01~\textrm{au}$ and
$\gamma_{\textrm{c}}=0.001~\textrm{au}$, respectively.}
\end{figure}

\begin{figure}
\includegraphics[width=0.65\textwidth]{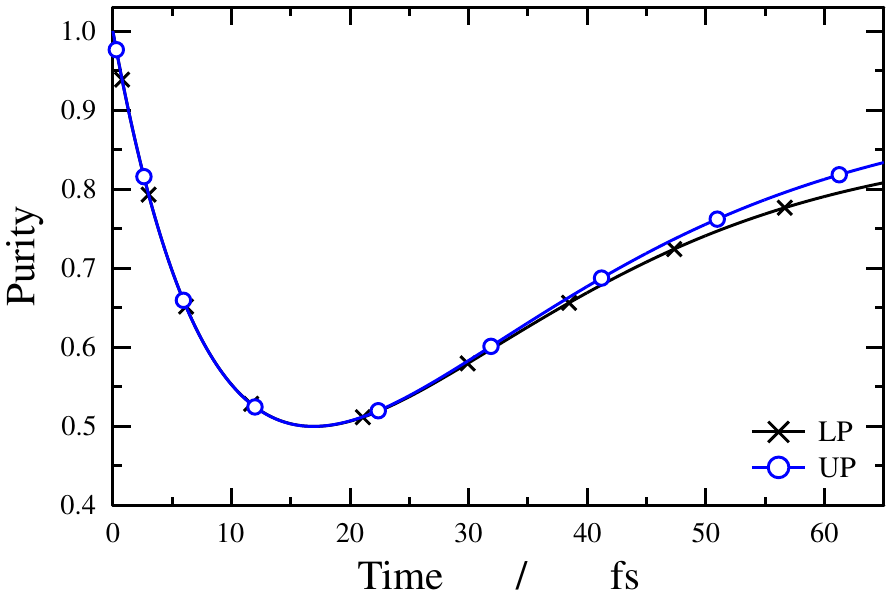}
\caption{\label{fig:special_purity}
Purity of the density matrix ($\textrm{tr}(\hat{\rho}^2)$) as a function of time for 
the initial state used in Figs. \ref{fig:special_1} and \ref{fig:special_2}.
The cavity wavenumber is set to $\omega_{\textrm{c}}=29957.2~\textrm{cm}^{-1}$ 
(labeled as LP) or $\omega_{\textrm{c}}=35744.8~\textrm{cm}^{-1}$ (UP).
The coupling strength and cavity decay rate equal $g=0.01~\textrm{au}$ and
$\gamma_{\textrm{c}}=0.001~\textrm{au}$, respectively.}
\end{figure}

This section presents results with a special initial state, that is,
the molecule is in the vibrational ground state of the ground electronic state and there is
1 photon in the cavity mode at $t=0$.
Fig. \ref{fig:special_1} compares populations of polaritonic states
and emission curves obtained with the Lindblad and non-Hermitian TDSE methods, showing good 
agreement for both cavity wavenumbers ($\omega_{\textrm{c}}=29957.2~\textrm{cm}^{-1}$ and 
$\omega_{\textrm{c}}=35744.8~\textrm{cm}^{-1}$). The same results are depicted in Fig.
\ref{fig:special_2} for the Lindblad and renormalized TDSE methods which show substantial
deviations for both cavity wavenumbers in this particular case. 
Finally, Fig. \ref{fig:special_purity} presents purity results.
In all figures, the coupling strength and cavity decay rate are set to 
$g=0.01~\textrm{au}$ and $\gamma_{\textrm{c}}=0.001~\textrm{au}$, respectively.

\clearpage

\bibliography{LindbladSchrodinger}

\end{document}